\documentclass [12pt,a4paper]{article}
\usepackage {amsmath}
\usepackage {graphicx}
\graphicspath{{images/}}
\usepackage {comment}
\usepackage{eqnarray}
\usepackage{relsize}
\usepackage {graphicx}
\graphicspath{{images/}}
\usepackage {comment}
\usepackage{authblk}
\usepackage{cite}
\usepackage{float}
\restylefloat{table}
\usepackage{multirow}  
\usepackage{amssymb}                    
\newcommand \bprime {\backprime\hspace{-.11em}      }   
\date{}
\title{Novel differential quadrature element method for higher order strain gradient elasticity theory}

\author{Md Ishaquddin\thanks{Corresponding author: \textit{E-mail address: ishaquddinmd@iisc.ac.in}}, S.Gopalakrishnan\thanks {\textit{E-mail address: krishnan@iisc.ac.in; Phone: +91-80-22932048}}}

\begin{document}

\maketitle
\vspace{-10mm}
\noindent\textit{Department of Aerospace Engineering,Indian Institute of Science 
Bengaluru 560012, India}\\

\begin{abstract}

In this paper, we propose a novel and efficient differential quadrature element based on Lagrange interpolation to solve a sixth order partial differential equations encountered in non-classical beam theories. These non-classical theories render displacement, slope and curvature as degrees of freedom for an Euler-Bernoulli beam. A generalize scheme is presented herein to implementation the multi-degrees degrees of freedom associated with these non-classical theories in a simplified and efficient way. The proposed element has displacement as the only degree of freedom in the domain, whereas, at the boundaries it has displacement, slope and curvature. Further, we extend this methodology and formulate two novel versions of plate element for gradient elasticity theory. In the first version, Lagrange interpolation is assumed in $x$ and $y$ directions and the second version is based on mixed interpolation, with Lagrange interpolation in $x$ direction and Hermite interpolation in $y$ direction. The procedure to compute the modified weighting coefficients by incorporating the classical and non-classical boundary conditions is explained. The efficiency of the proposed elements is demonstrated through numerical examples on static analysis of gradient elastic beams and plates for different boundary conditions. \\ \\
	\textbf{Keywords}: Differential quadrature element, gradient elasticity, sixth order pde, weighting coefficients, non-classical, Lagrange interpolation, Lagrange-Hermite mixed interpolation 

\end{abstract}
\section*{1.0 INTRODUCTION}

The differential quadrature method (DQM) is an efficient numerical tool for the solution of initial and boundary value problems. This technique was first introduced by Bellman et al.\cite{Bellman}. The major challenge in the application of DQM to structural problems is the implementation of multiple boundary conditions. To resolve this issue, many efficient and improved DQ schemes were developed in recent years. A comprehensive survey on the DQM development and the recent contributions to this field can be found in  \cite{BertR,Shu,Wangb}. Bert et al. proposed a $\delta$ - method to solve problems in structural mechanics \cite{Bert1,Bert2}. Subsequent developments in this technique led to the application of this methodology to variety of problems \cite{Bert3,Bert4,Bert5,Bert6,Bert7,Bert8}. As this method could not be generalized and had limitations on accuracy, an alternative scheme was proposed which accounted boundary conditions during the formulation of weighting coefficients \cite{Malik1}. However, this method also could not be generalized and worked only for few specific types of boundary conditions and structures with constrained regular edges. Further improvement in these schemes were discussed in the articles by Du et al. \cite{Du1,Du2}.  

To address the difficulties in applying the DQ method for structures with discontinuous loading and geometry, Striz et al. \cite{Striz1,Striz2} developed a quadrature element method (QEM), however, due to the use of $\delta$ - technique the scope of this method was limited. Later, Wang et al. \cite{Wang1} and Chen et al.\cite{Bert8} proposed a differential quadrature element method (DQEM) which assumes the slope as an independent degree of freedom at the boundary. The main advantage of this method is only one grid point is required to represent the multiple degrees of freedom at the boundary. Further improvement in this field led to the development of a new method called the generalized differential quadrature rule (GDQR) \cite{Wu1,Wu2,Wu3}. Following this many researchers applied DEQM and GDQR techniques to variety of structural problems \cite{Wang2,Wang3,Xing,Karami1,Karami2,Karami3,Wangold,Wangnew1}.

In the above DQEM and GDQR techniques, Hermit interpolation functions were used to determine the weighting coefficients. In contrast, Wang et al.\cite{Wangold} employed the weighting coefficients based on Lagrange interpolation functions. The research inclination in the aforesaid publications was towards the solution of fourth order partial differential equations which governs the problems related to classical beam and plate theories. The DQ solution for the sixth and eighth order differential equations using GDQR technique with Hermite interpolation function was reported by Wu et al. \cite{Six,Eigth}. They have demonstrated the capability for structural and fluid mechanics problems. Recently, Wang et al. \cite{Wangnew2} proposed a new differential quadrature element based on Hermite interpolation to solve a sixth order partial differential equation governing the non-local Euler-Bernoulli beam. In their study, they have computed the frequencies for various combination of boundary conditions. 

The classical continuum theories are effective for macro scale modelling of structural elements, neverthless, they lack efficiency to model the nano scale systems. These classical theories are governed by fourth order partial differential equations. To overcome this difficulty, several scale-dependent non-classical continuum theories are reported in the literature \cite{Mindlin1,Fleck,Mindlin2,Mindlin3,Koiter}. These non-classical continuum theories are enriched versions of classical continuum theories incorporating higher order terms in the constitutive relations. These higher order terms consists of stress and strain gradients accompanied with intrinsic length scale parameters which account for scale effects\cite{Lam1,Lam2,Reddy,Aifantis1}. One such class of gradient elasticity theory is the simplified theory by Mindlin et al.\cite{Mindlin1}, with one gradient elastic modulus and two classical constants for structural applications. This simplified theory was used by many researchers to study static, dynamic and buckling behaviour of gradient elastic beams \cite{Aifantis2,Aifantis3,Besko1b,Besko2b,Besko3b} and plates \cite{Besko1p,Besko2p,Besko3p}, by deriving the analytical solutions. Recently, Pegios et.al \cite{Pegios} developed a finite element model for gradient elastic Euler-Bernoulli beam and conducted static and stability analysis. The numerical solution of 2-D and 3-D gradient elastic structural problems using finite element and boundary element methods can be found in \cite{BEM}.

In this paper, we propose for the first time a novel differential quadrature beam element based on Lagrange interpolation to solve a sixth order partial differential equation associated with gradient elastic Euler-Bernoulli beam theory. Further, we extend this methodology and formulate two novel versions of plate element for gradient elastic Kirchhoff plate theory. In the first version, the Lagrange interpolations are used in both $x$ and $y$ direction, and later, mixed-interpolations are used with Lagrange interpolations in $x$ direction and $C^{2}$ continuous Hermite in $y$ direction. A novel way to impose the classical and non-classical boundary conditions for gradient elastic beam and plate elements are presented. A new procedure to compute the  higher order weighting coefficients for the proposed elements are explained in detail. The efficiency and the performance of the elements are established through numerical examples.

\section{Strain gradient elasticity theory}
Mindlin's \cite{Mindlin1} strain gradient micro-elasticity theory with two classical and one non-classical material constants is consider in the present study. The two classical material constants are Lam$e^{'}$ constants and the non-classical one is related to intrinsic bulk length $g$. In what follows, the theoretical basis required to formulate the differential quadrature beam and plate element for gradient elasticity theory are presented. Further, the classical and non-classical boundary conditions associated with the gradient elastic Euler-Bernoulli beam and Kirchhoff plate are discussed.

\subsection{Gradient elastic Euler-Bernoulli beam} \label{Section_Sg_beam}

The stress-strain relations for 1-D gradient elastic theory are defined as \cite{Vardo,Besko1b}
\begin{align*}
{\tau}&= 2\,\,\mu \,\, \varepsilon + \lambda \,\, {\text{tr}} \varepsilon \,\, \text{I} \\
{{\varsigma}}&= g^{2} \,\, [2 \,\, \mu \,\, \nabla\varepsilon + \lambda \,\, \nabla (\text{tr}\varepsilon) \,\, \text{I}] \tag{1}
\end{align*}

\noindent where, $\lambda$,$\,$ $\mu$ are Lam$e^{'}$ constants.$\nabla=\frac{\partial}{\partial x}+\frac{\partial}{\partial y}$ is the Laplacian operator and $\text{I}$ is the unit tensor. $\tau$, $\varsigma$ denotes Cauchy and higher order stress respectively, $\varepsilon$ and ($\text{tr}\,\varepsilon$) are the classical strain and its trace which are expressed in terms of displacement vector $\textit{w}$ as:
\begin{align*}
&{\varepsilon}= \frac{1}{2}(\nabla\textit{w}+\textit{w}\nabla)\,\,, \,\,\quad \text{tr}{\varepsilon}= \nabla\textit{w} \tag{2}
\end{align*}

It follows from the above equations the constitutive relations for an Euler-Bernoulli gradient beam can be stated as
\begin{align*}         
{\tau_{x}}= E\varepsilon_{x}, \quad \varsigma_{x}={g}^{2}\varepsilon_{x}^{'}, \quad \varepsilon_{x}=-z\dfrac{\partial^{2} w(x,t)}{\partial{x}^2}\tag{3}
\end{align*}

\noindent For the above state of stress and strain the strain energy expression in terms of displacement can be written as
\begin{align*}         
{U_{b}}= \frac{1}{2}\int_{0}^{L} EI\big[(w^{''})^{2}+g^{2}(w^{'''})^{2}\big]dx  \tag{4}
\end{align*}

The potential energy of the applied load is given by
\begin{align*}         
{V_{b}}= \int_{0}^{L}q_{b}(x)w{dx}+\big[Vw\big]_{0}^{L}-\big[M{w}^{'}\big]_{0}^{L}-\big[\bar{M}{w}^{''}\big]_{0}^{L}  \tag{5}
\end{align*}

The total potential energy of the beam is given by
\begin{align*}         
\Pi_{b}=U_{b}+V_{b}   \tag{6}
\end{align*}

\noindent  where, $E$, $A$ and $I$ are the Young's modulus, area, moment of inertia, respectively. $q_{b}$ and $w(x,t)$ are the transverse load and displacement of the beam. $V$, $M$ and $\bar{M}$ are shear force, bending moment and higher order moment acting on the beam.

Using the principle of minimum potential energy \cite{Reddyb}:
\begin{align*}         
\delta\Pi_{b}=\delta(U_{b}+V_{b})=0   \tag{7}
\end{align*}

\noindent and performing integration by parts, we get the governing equation for a gradient elastic Euler-Bernoulli beam as
\begin{align*} \label{eq:EOM_Beam}        
EI(w^{\prime\bprime\prime}-{g}^{2}w^{\bprime\prime\prime})+q_{b}=0 \tag{8}
\end{align*}

\noindent and the associated boundary conditions are: \\

\noindent \textit{Classical} :     
\begin{align*}  \label{eq:BC_Cl_Beam}     
V&=EI[w^{'''}-{g}^{2}w^{\bprime\prime}]=0 \hspace{0.5cm}\text{or}\hspace{0.5cm} w=0,\hspace{0.5cm}\text{at}\,\,x=(0,L)\\
 M&=EI[w^{''}-{g}^{2}w^{\prime\bprime\prime}]=0 \hspace{0.5cm}\text{or} \hspace{0.5cm}w^{'}=0,\hspace{0.5cm}\text{at}\,\,x=(0,L)
\tag{9}
\end{align*}

\noindent \textit { Non-classical} :         
\begin{align*}  \label{eq:BC_NCl_Beam}       
\bar{M}&=[{g}^{2}EIw^{'''}]=0 \hspace{0.5cm}\text{or}\hspace{0.5cm} w^{''}=0,\,\,\,\text{at}\,\,x=(0,L)		\tag{10}
\end{align*} \\
\noindent The list of classical and non-classical boundary conditions employed in the present study for a gradient elastic Euler-Bernoulli beam are as follows \\
 
 \noindent \text{Simply supported} :\\
\noindent \textit{classical} :\,\,$w=M=0$ , \,\,\,\textit{non-classical} : $w^{''}=0$ \,\,at $x=(0,L)$ \\ 

\noindent \text{Clamped} :\\
\noindent \textit{classical} :\,\,$w=w^{'}=0$ , \,\,\,\textit{non-classical} : $w^{''}=0$ \,\,at $x=(0,L)$ \\

Next, the constitutive relations, governing equation and the associated classical and non-classical boundary conditions for a gradient Kirchhoff plate are presented.

\subsection{Gradient elastic Kirchhoff plate} \label{Section_Sg_plate}

The strain-displacement relations for a Kirchhoff's plate theory can be defined as \cite{Timo}
\begin{align*}
{\varepsilon_{xx}}= -z{\bar{w}}_{xx}, \quad {\varepsilon_{yy}}=-z\bar{w}_{yy} , \quad \gamma_{xy}=2\varepsilon_{xy}=-2z{\bar{w}}_{xy} \tag{11}
\end{align*}

\noindent where, $\bar{w}(x,y,t)$  is transverse displacement of the plate. 
The stress-strain relations for a gradient elastic Kirchhoff plate is given by \cite{Vardo,Koiter}:\\

\noindent \textit{Classical:}
\begin{align*}
{\tau}_{xx}= &\frac{E}{1-\nu^{2}}(\varepsilon_{xx}+\nu\varepsilon_{yy})\\
{\tau}_{yy}= &\frac{E}{1-\nu^{2}}(\varepsilon_{yy}+\nu\varepsilon_{xx})\tag{12} \\
{\tau}_{xy}= &\frac{E}{1+\nu}\varepsilon_{xy}  
\end{align*}
\noindent \textit{Non-classical:}
\begin{align*}
{\varsigma_{xx}}= &g^{2} \frac{E}{1-\nu^{2}} \nabla^{2}(\varepsilon_{xx}+\nu\varepsilon_{yy})\\
{\varsigma_{yy}}= &g^{2} \frac{E}{1-\nu^{2}}\nabla^{2}(\varepsilon_{yy}+\nu\varepsilon_{xx})\tag{13} \\
{\varsigma_{xy}}= &g^{2} \frac{E}{1+\nu}\nabla^{2}\varepsilon_{xy} 
\end{align*}

\noindent where,  $\tau_{xx}$, $\tau_{yy}$,$\tau_{xy}$, are the classical Cauchy stresses and $\varsigma_{xx}$,$\varsigma_{yy}$, $\varsigma_{xy}$ denotes higher order stresses related to gradient elasticity.

 The strain energy for a gradient elastic Kirchhoff plate is given by \cite{Koiter,Besko3p}:
\begin{align*}        
						U_{p}=U_{cl}+U_{sg} \tag{14}
\end{align*}

\noindent where, $U_{cl}$ and $U_{sg}$ are the classical and gradient elastic strain energy given by
\begin{align*}        
{U}_{cl}= \frac{1}{2}D\,\int\int_{A}\Big[\bar{w}_{xx}^{2}+\bar{w}_{yy}^{2}+2\bar{w}_{xy}^{2}+2\,\nu\,(\bar{w}_{xx}\bar{w}_{yy}
-\bar{w}_{xy}^{2})\Big]dxdy \tag{15}
\end{align*}

\begin{align*}        
{U}_{sg}= &\frac{1}{2}g^{2}D\,\int\int_{A}\Big[\bar{w}_{xxx}^{2}+\bar{w}_{yyy}^{2}+3(\bar{w}_{xyy}^{2}+
\bar{w}_{xxy}^{2})+ 2\,\nu\,(\bar{w}_{xyy}\bar{w}_{xxx}+\\& \hspace{2.0cm}\,\,\,\, \bar{w}_{xxy}\bar{w}_{yyy}-\bar{w}_{xyy}^{2}-\bar{w}_{xxy}^{2}\Big]dxdy \tag{16}
\end{align*}
 	 	
\noindent here, $D=\frac{E{h}^3}{12(1-\nu^{2})}$.\\

The potential energy of the external load is defined as
\begin{align*}         
{W_{p}}= \int\int_{A}q_{0}(x,y)\bar{w}\,{dx}{dy}+\int_{s}V_{n}\bar{w}\,{ds}-\int_{s}M_{n}\,\bar{w}^{'}(n)\,ds-\int_{s}\bar{M}_{n}\,\bar{w}^{''}(n)\,{ds}  \tag{17}
\end{align*}
 	 	
\noindent where $q_{0}$ is the transverse load on the plate. $n$ and $s$ are the normal and tangential to the boundary point corresponding to $x$ and $y$ coordinates axis. \\

Using the principle of minimum potential energy and performing integration by parts over the area, we obtain the governing equation for a gradient elastic Kirchhoff plate as

\begin{align*} \label{eq:EOM_Plate}        
D\nabla^{4}\bar{w}-g^{2}D\nabla^{6}\bar{w}-q_{0}=0\tag{18} 
\end{align*}
where,
\begin{align*}         
&\nabla^{4}=\frac{\partial^{4} \bar{w}}{\partial x^{4}}+\frac{\partial^{4} \bar{w}}{\partial y^{4}}+2\frac{\partial^{4} \bar{w}}{\partial x^{2}\partial y^{2}},\\ 
&\nabla^{6}=\frac{\partial^{6} \bar{w}}{\partial x^{6}}+\frac{\partial^{6} \bar{w}}{\partial y^{6}}+3\frac{\partial^{6} \bar{w}}{\partial x^{4}\partial y^{2}}+3\frac{\partial^{6} \bar{w}}{\partial x^{2}\partial y^{4}}
\end{align*}

\noindent the associated boundary conditions for the plate with domain defined over ($0\leq x \leq l_{x}$), ($0\leq y \leq l_{y}$), are listed below.\\   

\noindent \textit{Classical boundary conditions} :
\begin{align*}    
 V_{x}=-D\bigg(\frac{\partial^{3} \bar{w}}{\partial x^{3}}+(2-\nu) \frac{\partial^{3} \bar{w}}{\partial x\partial y^{2}}\bigg)+g^{2}D\bigg[ \frac{\partial^{5} \bar{w}}{\partial x^{5}}+(3-\nu) \frac{\partial^{5} \bar{w}}{\partial x\partial y^{4}}+3 \frac{\partial^{5} \bar{w}}{\partial y^{2}\partial x^{3}}\bigg]=0 \\ \text{or} \\ \bar{w}=0,\,\, \text{at}\,\,x=(0,l_{x})
\end{align*}
\begin{align*} \label{eq:BC_Cl_Plate_V}   
V_{y}=-D\bigg(\frac{\partial^{3} \bar{w}}{\partial y^{3}}+(2-\nu) \frac{\partial^{3} \bar{w}}{\partial y\partial x^{2}}\bigg)+g^{2}D\bigg[ \frac{\partial^{5} \bar{w}}{\partial y^{5}}+(3-\nu) \frac{\partial^{5} \bar{w}}{\partial y\partial x^{4}}+3 \frac{\partial^{5} \bar{w}}{\partial x^{2}\partial y^{3}}\bigg]=0 \\ \text{or} \\ \bar{w}=0,
\,\,\text{at}\,\,y=(0,l_{y})   \tag{19}
\end{align*}

\begin{align*} \label{eq:BC_Cl_Plate_M}   
M_{x}=-D\bigg(\frac{\partial^{2} \bar{w}}{\partial x^{2}}+\nu \frac{\partial^{2} \bar{w}}{\partial y^{2}}\bigg)+g^{2}D\bigg[ \frac{\partial^{4} \bar{w}}{\partial x^{4}}+\nu \frac{\partial^{4} \bar{w}}{\partial y^{4}}+(3-\nu) \frac{\partial^{4} \bar{w}}{\partial x^{2}\partial y^{2}}\bigg]=0 \\\text{or}\\\hspace{0.5cm} \bar{w}_{x}=0,\,\,\text{at}\,\,x=(0,l_{x})\\  
M_{y}=-D\bigg(\frac{\partial^{2} \bar{w}}{\partial y^{2}}+\nu \frac{\partial^{2} \bar{w}}{\partial x^{2}}\bigg)+g^{2}D\bigg[ \frac{\partial^{4} \bar{w}}{\partial y^{4}}+\nu \frac{\partial^{4} \bar{w}}{\partial x^{4}}+(3-\nu) \frac{\partial^{4} \bar{w}}{\partial x^{2}\partial y^{2}}\bigg]=0 \\\text{or} \\ \hspace{0.5cm} \bar{w}_{y}=0,\,\,\text{at}\,\,y=(0,l_{y})\\  
\tag{20}
\end{align*}

\noindent \textit { Non-classical boundary conditions} :  
\begin{align*} \label{eq:BC_NCl_Plate}        
\bar{M}_{x}=-g^{2}D\bigg(\frac{\partial^{3} \bar{w}}{\partial x^{3}}+\nu \frac{\partial^{3} \bar{w}}{\partial x\partial y^{2}}\bigg)=0 \,\,\text{or}\,\,\, \bar{w}_{xx}=0,\,\,\,\text{at}\,\,x=(0,l_{x})	\\	
\bar{M}_{y}=-g^{2}D\bigg(\frac{\partial^{3} \bar{w}}{\partial y^{3}}+\nu \frac{\partial^{3} \bar{w}}{\partial y\partial x^{2}}\bigg)=0 \,\,\text{or}\,\,\, \bar{w}_{yy}=0,\,\,\,\text{at}\,\,y=(0,l_{y}) \tag{21}
\end{align*}

\noindent The concentrated force at the free corner is given by
\begin{align*} \label{eq:BC_NCl_Plate_CForce}         
&R=2D(1-\nu)\frac{\partial^2 \bar{w}}{\partial x\partial y}-2g^{2}D(1-\nu)\bigg[\frac{\partial^4 \bar{w}}{\partial x^3\partial y}+\frac{\partial^4 \bar{w}}{\partial y^3\partial x}\bigg]	\tag{22}
\end{align*}

\noindent  where $l_{x}$ and $l_{y}$ are the length and width of the plate. $V_{x}$,\,$V_{y}$ are the shear force, $M_{x}$,\,$M_{y}$ are the bending moment and $\bar{M}_{x}$,\,$\bar{M}_{y}$ are the higher order moment. $R$ is the concentrated force at the free corner.\\

\noindent The different boundary conditions employed in the present study for a gradient elastic Kirchhoff plate are:\\

\noindent \textit{Simply supported edge} :\\
\noindent $\bar{w}=M_{x}=\bar{w}_{x}=0$ \,or \,\,$\bar{w}_{xx}=0$ \hspace{0.2cm}at $x=0, l_{x}$ \\
\noindent $\bar{w}=M_{y}=\bar{w}_{y}=0$ \,or \,\,$\bar{w}_{yy}=0$\hspace{0.35cm}at $y=0, l_{y}$ \\

\noindent \textit{Clamped edge} :\\
\noindent $\bar{w}=\bar{w}_{x}=\bar{w}_{xx}=0$ \hspace{0.2cm}at $x=0, l_{x}$\\
\noindent $\bar{w}=\bar{w}_{y}=\bar{w}_{yy}=0$ \hspace{0.2cm}at $y=0, l_{y}$ \\

\noindent \textit{Free edge} :\\
\noindent $V_{x}=M_{x}=\bar{M}_{x}=0 $  \hspace{0.2cm}at $x=0, l_{x}$ \\
\noindent $V_{y}=M_{y}=\bar{M}_{y}=0 $  \hspace{0.2cm}at $y=0, l_{y}$\\

\noindent \textit{Free corner} :\\
\noindent $R=M_{x}=M_{y}=\bar{M}_{x}=\bar{M}_{y}=0$

\section{Differential quadrature elements for gradient elasticity theory}

In this section, first, we formulate a differential quadrature element based on Lagrangian interpolation function for 1-D gradient elastic Euler-Bernoulli beam. Next, we develop two new versions of gradient plate element with different choice of interpolation functions. The grid employed in the present study is unequal Gauss\textendash Lobatto\textendash Chebyshev points given by
\begin{align*}    
z_{i}=\frac{1}{2}\Bigg[1-cos{\frac{(i-1)\pi}{N-1}} \Bigg] \tag{23}
\end{align*}

\noindent where $N$ is the number of grid points and $z$ are the coordinates of the grid. For the plate analysis $N=N_{x}=N_{y}$ is employed.

\subsection{Differential quadrature element for gradient Euler-Bernoulli beam} \label{Lagrange_Beam_section}

The $n$th order derivative of the deflection $w(x,t)$ at location $x_{i}$ for a N-node 1-D beam element is assumed as
\begin{align*}    
w_{i}^{n}(x,t)=\sum_{j=1}^{N} L_{j}^{n}(x)w_{j} \tag{24}
\end{align*}

\noindent $L_{j}(x)$ are Lagrangian interpolation functions in $x$ co-ordinate. The Lagrange interpolation functions can be defined as\cite{Wangb,Shu},
\begin{align*}    
L_{j}(x)=\frac{\beta(x)}{\beta(x_{j})}=\prod_{\substack{k=1 \\ (k\neq j)}}^{N}\frac{(x-x_{k})}{(x_{j}-x_{k})}   \tag{25}
\end{align*}

\noindent where \\
$\beta(x)=(x-x_{1})(x-x_{2})\cdots(x-x_{j-1})(x-x_{j+1})\cdots(x-x_{N})$ \\
$\beta(x_{j})=(x_{j}-x_{1})(x_{j}-x_{2})\cdots(x_{j}-x_{j-1})(x_{j}-x_{j+1})\cdots)(x_{j}-x_{N})$

The first order derivative of the above shape functions can be written as
\begin{align*}    
A_{ij}={L}^{'}_{j}(x_{i})\begin{cases}
\mathlarger\prod_{\substack{k=1 \\ (k\neq i,j)}}^{N}(x_{i}-x_{k})/\mathlarger\prod_{\substack{k=1 \\ (k\neq j)}}^{N}=(x_{j}-x_{k})\,\,\,\, (i\neq j)\\ \\ 
\mathlarger{\sum}_{\substack{k=1 \\ (k\neq i)}}^{N}\frac{1}{(x_{i}-x_{k})}
\end{cases}\tag{26}  
\end{align*}

The conventional higher order weighting coefficients are computed as
\begin{align*}    
B_{ij}=\sum_{k=1}^{N} A_{ik}A_{kj} \,,\quad
C_{ij}=\sum_{k=1}^{N} B_{ik}A_{kj} \,,\quad 
D_{ij}=\sum_{k=1}^{N} B_{ik}B_{kj}\,,\quad(i,j=1,2,...,N)\tag{27}
\end{align*}
here, $B_{ij}$ , $C_{ij}$ and $D_{ij}$ are weighting coefficients for second, third, and fourth order derivative, respectively.\\

\begin{figure}[H]
\includegraphics[width=1.0\textwidth]{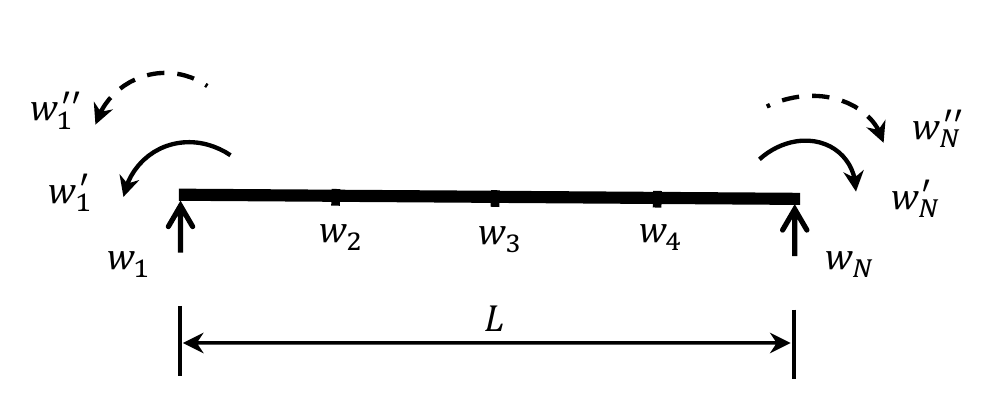}
\centering
\caption{A typical differential quadrature element for an Euler-Bernoulli gradient beam.}
\label{fig:beam}
\end{figure}

Let us consider a N-node gradient Euler-Bernoulli beam element as shown in the Figure \ref{fig:beam}.

Each interior node has displacement $w$ as the only degree of freedom, and the boundary nodes has 3 degrees of freedom $w$, $w^{'}$, $w^{''}$. These extra boundary degrees of freedom related to slope and curvature are introduced in to the formulation through modifying the conventional weighting coefficients. The new displacement vector now includes the slope and curvature as additional degrees of freedom at the element boundaries as: $w^{b}=\{w_{1},\cdots w_{N},w^{'}_{1},w^{'}_{N},w^{''}_{1},w^{''}_{N}\}$. The modified weighting coefficient matrices accounting for slope and curvature degrees of freedom at the boundaries are derived as follows:\\ 

\noindent \textit{First order derivative matrix}: 
\begin{align*} \label{eq:WC_Beam_Aij}   
\bar{A}_{ij}=\begin{cases}
A_{ij} \,\,\,\, (i,j=1,2,\cdots,N)\\ \\
0 \,\,\,\,\,\,\,(i=1,2,\cdots,N;\,j=N+1,\cdots,N+4)\tag{28}
\end{cases} 
\end{align*}
\noindent \textit{Second order derivative matrix}: 
\begin{align*} \label{eq:WC_Beam_B2ij}
\bar{B}_{ij}=\begin{cases}
B_{ij} \,\,\,\, (i=2,3,\cdots,N-1; \,\,j=1,2,\cdots,N)\\ \\
0 \,\,\,\,\,\,\,(i=2,3,\cdots,N-1; \,\, j=N+1,\cdots,N+4)\tag{29}
\end{cases} 
\end{align*}
\begin{align*} \label{eq:WC_Beam_B2ij}
\bar{B}_{ij}=\sum_{k=2}^{N-1}A_{ik}A_{kj} \,\,\, (i=1,N; \,\, j=1,2,\cdots,N)\tag{30}
\end{align*} 
\begin{align*} \label{eq:WC_Beam_B2ij}
\bar{B}_{i(N+1)}=A_{i1}\,\,; \,\, \, \bar{B}_{i(N+2)}=A_{iN} \,\,\,\,(i=1,N)\tag{31}
\end{align*} 

\noindent \textit{Third order derivative matrix}: 
\begin{align*}\label{eq:WC_Beam_C1ij}
\bar{C}_{ij}=\begin{cases}
\mathlarger{\sum}_{j=1}^{N}\mathlarger\sum_{k=1}^{N}\bar{B}_{ik}A_{kj} \,\,\, (i=2,3,...,N-1) \\ \\ 
0 \,\,\,\,\,\,\,\,\,\, (i=2,3,\cdots,N-1; \,\, j=N+1,\cdots,N+4)\tag{32}
\end{cases}
\end{align*} 
\begin{align*} \label{eq:WC_Beam_C2ij2}
\bar{C}_{ij}=\sum_{k=2}^{N-1}B_{ik}A_{kj}\,\,\  (i=1,N; \,\,j=1,2,\cdots,N)\tag{33}
\end{align*}
\begin{align*} \label{eq:WC_Beam_C2ij22}
\bar{C}_{i(N+3)}=A_{i1}\,\,; \,\, \, \bar{C}_{i(N+4)}=A_{iN} \,\,\,\,(i=1,N)\tag{34}
\end{align*}

\noindent \textit{Fourth order derivative matrix}: 
\begin{align*}\label{eq:WC_Beam_Eij}
\bar{D}_{ij}=\sum_{j=1}^{N+4}\sum_{k=1}^{N} B_{ik}\bar{B}_{kj}  \,\,\,\,\,\,\, (i=1,2,...,N)\tag{35}
\end{align*} 

\noindent \textit{Fifth order derivative matrix}: 

\begin{align*} \label{eq:WC_Beam_D1ij}
{V}_{ij}=\begin{cases}
\bar{D}_{ij} \,\,\,\, (i=2,3,\cdots,N-1; \,\, j=1,2,\cdots,N)\\ \\
0 \,\,\,\,\,\,\,(i=2,3,\cdots,N-1; \,\, j=N+1,\cdots,N+4)\tag{36}
\end{cases} 
\end{align*}
\begin{align*} \label{eq:WC_Beam_D2ij}
{V}_{ij}=\sum_{k=2}^{N-1}B_{ik}B_{kj} \,\,\, (i=1,N; \,\,\,\, j=1,2,\cdots,N)\\ 
{V}_{i(N+3)}=B_{i1}\,\,; \,\,\, \, V_{i(N+4)}=B_{iN} \,\,\,(i=1,N)\tag{37}
\end{align*} 
\begin{align*}\label{eq:WC_Beam_Eij}
\bar{E}_{ij}=\sum_{k=1}^{N} A_{ik}{V}_{kj}  \,\,\,\,\,(i=1,2,...,N; \,\,j=1,2,...,N+4)\tag{38}
\end{align*} 

\noindent \textit{Sixth order derivative matrix}: 
\begin{align*} \label{eq:WC_Beam_Fij}
\bar{F}_{ij}=\sum_{k=1}^{N} B_{ik}{V}_{kj}  \,\,\,\,\,(i=1,2,...,N; \,\,j=1,2,...,N+4)\tag{39}
\end{align*}

Here, $\bar{A}_{ij}$, $\bar{B}_{ij}$, $\bar{C}_{ij}$, $\bar{D}_{ij}$, $\bar{E}_{ij}$ and $\bar{F}_{ij}$ are first to sixth order modified weighting coefficients matrices, respectively. Using the above Equations (\ref{eq:WC_Beam_Aij})-(\ref{eq:WC_Beam_Fij}), the governing differential equation (\ref{eq:EOM_Beam}), in terms of the differential quadrature at inner grid points is written as
 \begin{align*}         
	 EI\sum_{j=1}^{N+4}\bar{D}_{ij}w^{b}_{j}-g^{2}EI\sum_{j=1}^{N+4}\bar{F}_{ij}w^{b}_{j}= q_{b}(x_{i}) \,\,\,\,\,\, (i=2,3,...,N-1)\tag{40}
\end{align*} 
	 
The boundary forces given by Equations (\ref{eq:BC_Cl_Beam})-(\ref{eq:BC_NCl_Beam}), in terms of differential quadrature are expressed as \\

\noindent Shear force:
 \begin{align*} \label{eq:Boundary_forces_Beam_V}        
	 V_{i}=EI\sum_{j=1}^{N+4}\bar{C}_{ij}w^{b}_{j}-g^{2}EI\sum_{j=1}^{N+4}\bar{E}_{ij}w^{b}_{j} \,\,\,\,\,\, (i=1,N)\tag{41}
\end{align*} 
\noindent Bending moment:
 \begin{align*} \label{eq:Boundary_forces_Beam_M}        
	 M_{i}=EI\sum_{j=1}^{N+4}\bar{B}_{ij}w^{b}_{j}-g^{2}EI\sum_{j=1}^{N+4}\bar{D}_{ij}w^{b}_{j} \,\,\,\,\,\, (i=1,N)\tag{42}
\end{align*} 
\noindent Higher order moment:
 \begin{align*} \label{eq:Boundary_forces_Beam_MM}        
	 \bar{M}_{i}=g^{2}EI\sum_{j=1}^{N+4}\bar{C}_{ij}w^{b}_{j} \,\,\,\,\,\ (i=1,N)\tag{43}
\end{align*}  

\noindent here $i=1$ and $i=N$ correspond to the left support $x=0$ and right support $x=L$ of the beam, respectively.

Once the boundary conditions in Equations (\ref{eq:BC_Cl_Beam})-(\ref{eq:BC_NCl_Beam}) are applied, we get the following system of equations in the matrix form as
\begin{align*}\label{eq:Boundary_disp_Beam}
\begin{bmatrix}
\,k_{bb} & \phantom{-}k_{bd} \\ \\ 
\,k_{db} & \phantom{-}k_{dd} \\ \\ 
 \end{bmatrix}\begin{Bmatrix}
\,\Delta_{b} \\ \\ 
\,\Delta_{d}  \\ \\ 
 \end{Bmatrix}=\begin{Bmatrix}
\,f_{b} \\ \\ 
\,f_{d}  \\ \\ 
 \end{Bmatrix} \tag{44}
\end{align*}

\noindent where the subscript $b$ and $d$ indicates the boundary and domain of the beam. $f_{b}$, $\Delta_{b}$ and $f_{d}$, $\Delta_{d}$ are the boundary and domain forces and displacements of the beam, respectively. Now expressing  the system of equations in terms of domain dofs $\Delta_{d}$, we get
\begin{align*} \label{eq:Domain_disp_Beam}
\Big[k_{dd}-k_{db}k_{bb}^{-1}k_{bd}\Big] {\Big\{}\Delta_{d}{\Big\}}=
{\Big\{}f_{d}-k_{db}k_{bb}^{-1}f_{b}{\Big\}}\tag{45}
\end{align*}
The solution of the above system of equations renders the displacements at the domain nodes of the beam element. The boundary displacements are computed from Equation (\ref{eq:Boundary_disp_Beam}), and forces are computed from the Equations (\ref{eq:Boundary_forces_Beam_V})-(\ref{eq:Boundary_forces_Beam_MM}).

\subsection{Differential quadrature element for gradient elastic plates}

Here, we present two versions of novel differential quadrature element for a gradient elastic Kirchhoff plate. First, the differential quadrature element based on Lagrange interpolation in $x$ and $y$ direction is formulated. Next, the differential quadrature element based on Lagrange-Hermite mixed interpolation, with Lagrangian interpolation is $x$ direction and Hermite interpolation assumed in $y$ direction is presented. Similar to the beam elements discussed in the previous section, the plate element has only displacement $\bar{w}$ as degrees of freedom in the domain and at the plate edges it has 3 degrees of freedom $\bar{w}$, $\bar{w}_{x}$, $\bar{w}_{xx}$ or $\bar{w}_{y}$, $\bar{w}_{yy}$ depending upon the edge. At the corners the element has five degrees of freedom $\bar{w}$, $\bar{w}_{x}$, $\bar{w}_{y}$, $\bar{w}_{xx}$ and $\bar{w}_{yy}$. The new displacement vector now includes the slope and curvature as additional degrees of freedom at the element boundaries given by: $\tilde{w}=\{\bar{w}_{i},\cdots,\bar{w}_{N\times N},\bar{w}^{j}_{x},\cdots,\bar{w}^{j}_{y},\cdots,\bar{w}^{j}_{xx},\cdots,\bar{w}^{j}_{yy},\cdots\}$, where $(i=1,2,\cdots,N\times N; \,\, j=1,2,\cdots,4N)$.  
\begin{figure}[H]
\includegraphics[width=1.0\textwidth]{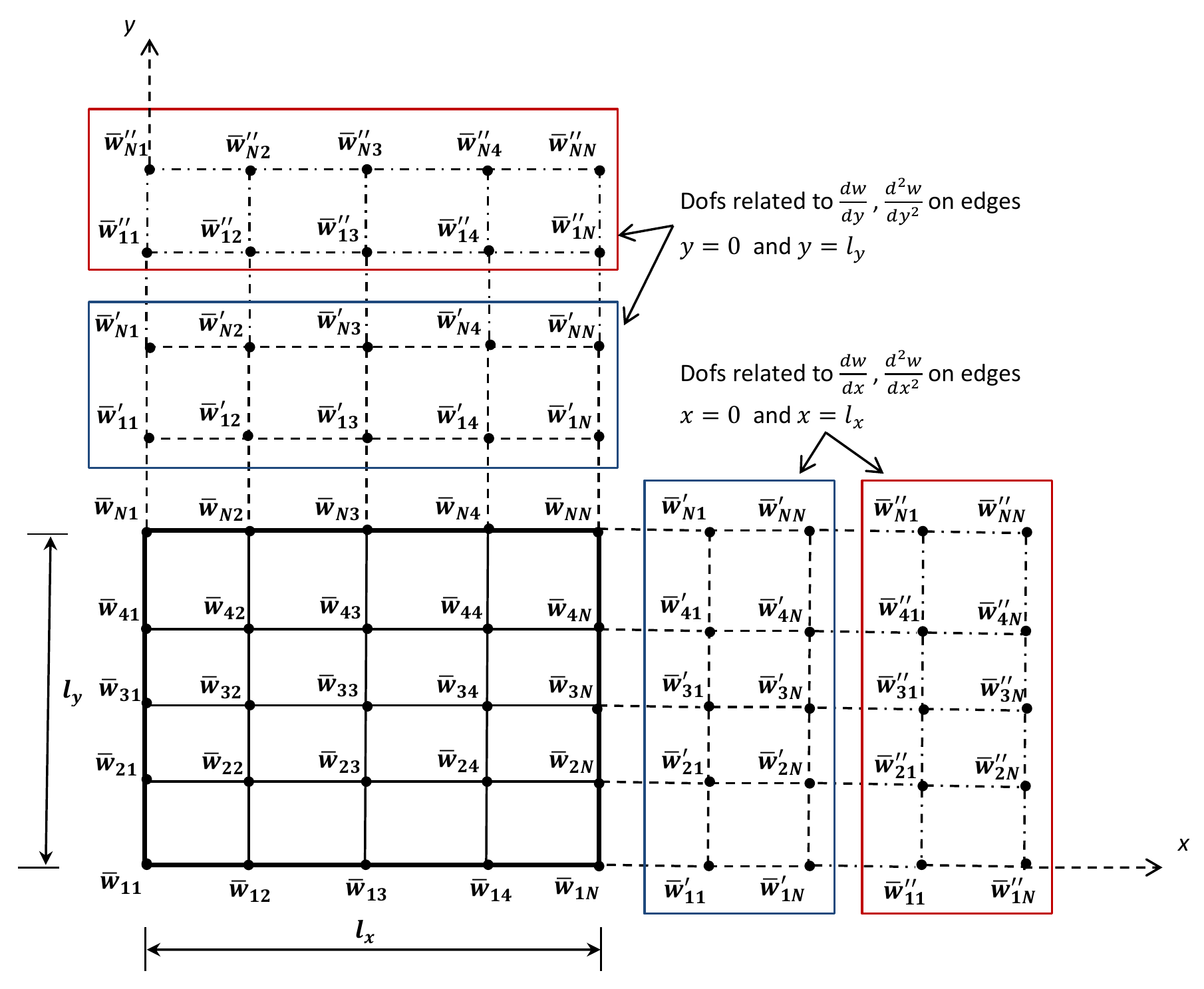}
\centering
\caption{A typical differential quadrature element for a gradient elastic Kirchhoff plate with $N=N_{x}=N_{y}=5$.}
\label{fig:Plate}
\end{figure}
 A differential quadrature gradient plate element for a $N_{x} \times N_{y}$ grid is shown in the Figure \ref{fig:Plate}. Here, $N_{x}=N_{y}=5$ are the number of grid points in $x$ and $y$ directions, respectively. It can be seen that the element has three degrees of freedom on each edge and five degrees of freedom at the corners. In the figure, the slope $\bar{w}^{'}$ and curvature $\bar{w}^{''}$ dofs with first subscript $1$ and $N$, correspond to the edges $y=0$ and $y=l_{y}$, respectively. Similarly, the slope $\bar{w}^{'}$ and curvature $\bar{w}^{''}$ dofs with second subscript $1$ and $N$, correspond to the edges $x=0$ and $x=l_{x}$, respectively. 

\subsubsection{Lagrange interpolation based differential quadrature element for gradient elastic plates}
The deflection for a $N_{x} \times N_{y}$ node differential quadrature rectangular plate element is assumed as
\begin{align*}    
w(x,y,t)=\sum_{p=1}^{N_{x}} \sum_{q=1}^{N_{y}}L_{p}(x)L_{q}(y)\tilde{w}_{pq}(t) \tag{46}
\end{align*}
\noindent where $\tilde{w}_{pq}(t)$ is the nodal deflection vector and $\bar{L}_{p}(x)$ and $\bar{L}_{q}(y)$ are the Lagrange interpolation functions in $x$ and $y$ directions, respectively. The slope and curvature degrees of freedom at the element boundaries are accounted while computing the weighting coefficients of higher order derivatives as discussed in section \ref{Lagrange_Beam_section}. Using the 1-D Lagrange interpolation functions derived in the section \ref{Lagrange_Beam_section}, the governing differential equation (\ref{eq:EOM_Plate}) in the differential quadrature syntax at inner grid points can be expressed as
 \begin{align*}         
&D\Bigg[\sum_{r=1}^{N_{x}+4}\bar{D}_{pr}^{x}\tilde{w}_{rs}+
 2\sum_{q=1}^{N_{x}+4}\sum_{r=1}^{N_{y}+4}\bar{B}_{pq}^{x}
 \bar{B}_{sr}^{y}\tilde{w}_{qr}+
 \sum_{r=1}^{N_{y}+4}\bar{D}_{sr}^{y}\tilde{w}_{pr}\Bigg]-\\
 & g^{2}D\Bigg[\sum_{r=1}^{N_{x}+4}\bar{F}_{pr}^{x}\tilde{w}_{rs}+  
 3\sum_{q=1}^{N_{x}+4}\sum_{r=1}^{N_{y}+4}\bar{D}_{pq}^{x}
 \bar{B}_{sr}^{y}\tilde{w}_{qr}+
  3\sum_{q=1}^{N_{x}+4}\sum_{r=1}^{N_{y}+4}\bar{B}_{pq}^{x}
 \bar{D}_{sr}^{y}\tilde{w}_{qr}+\\ &\hspace{3cm}
 \sum_{r=1}^{N_{y}+4}\bar{F}_{sr}^{y}\tilde{w}_{pr}\Bigg]= \bar{q}_{o}(x_{p},y_{s}) \\ \,\,\,\ &\hspace{5cm}(p=2,3,...,N_{x}-1; \,\,s=2,3,...,N_{y}-1)   \tag{47}
\end{align*} 
	 
The boundary forces, Equation (\ref{eq:BC_Cl_Plate_M})-(\ref{eq:BC_NCl_Plate_CForce}), in terms of differential quadrature form are expressed as \\
\noindent Shear force:
 \begin{align*}    \label{Plate_SFx_dqm}      
V_{x}=&-D\Bigg(\sum_{r=1}^{N_{x}+4}\bar{C}_{pr}^{x}\tilde{w}_{rs}+
(2-\nu)\sum_{q=1}^{N_{x}+4}\sum_{r=1}^{N_{y}+4}\bar{A}_{pq}^{x}
 \bar{B}_{sr}^{y}\tilde{w}_{qr}\Bigg) +\\
 & g^{2}D\Bigg[\sum_{r=1}^{N_{x}+4}\bar{E}_{pr}^{x}\tilde{w}_{rs}+
(3-\nu)\sum_{q=1}^{N_{x}+4}\sum_{r=1}^{N_{y}+4}\bar{A}_{pq}^{x}
 \bar{D}_{sr}^{y}\tilde{w}_{qr}+
3\sum_{q=1}^{N_{x}+4}\sum_{r=1}^{N_{y}+4}\bar{C}_{pq}^{x}
 \bar{B}_{sr}^{y}\tilde{w}_{qr}\Bigg]
 \\ \,\,\,\ &\hspace{5cm}(p=1,N_{x}; \,\,s=2,3,...,N_{y}-1)   \tag{48}
\end{align*} 
 \begin{align*}   \label{Plate_SFy_dqm}       
V_{y}=&-D\Bigg(\sum_{r=1}^{N_{y}+4}\bar{C}_{sr}^{y}\tilde{w}_{pr}+
(2-\nu)\sum_{q=1}^{N_{x}+4}\sum_{r=1}^{N_{y}+4}\bar{B}_{pq}^{x}
 \bar{A}_{sr}^{y}\tilde{w}_{qr}\Bigg) +\\
 & g^{2}D\Bigg[\sum_{r=1}^{N_{y}+4}\bar{E}_{sr}^{y}\tilde{w}_{pr}+
(3-\nu)\sum_{q=1}^{N_{x}+4}\sum_{r=1}^{N_{y}+4}\bar{D}_{pq}^{x}
 \bar{A}_{sr}^{y}\tilde{w}_{qr}+
3\sum_{q=1}^{N_{x}+4}\sum_{r=1}^{N_{y}+4}\bar{B}_{pq}^{x}
 \bar{C}_{sr}^{y}\tilde{w}_{qr}\Bigg]
 \\ \,\,\,\ &\hspace{5cm}(p=2,3,...,N_{x}-1; \,\,s=1,N_{y})   \tag{49}
\end{align*} 

\noindent Bending moment:
 \begin{align*}   \label{Plate_BMx_dqm}      
M_{x}=&-D\Bigg(\sum_{r=1}^{N_{x}+4}\bar{B}_{pr}^{x}\tilde{w}_{rs}+
\nu \sum_{r=1}^{N_{y}+4}\bar{B}_{sr}^{y}\tilde{w}_{pr}\Bigg) +\\
 & g^{2}D\Bigg[\sum_{r=1}^{N_{x}+4}\bar{D}_{pr}^{x}\tilde{w}_{rs}+
\nu\sum_{r=1}^{N_{y}+4}\bar{D}_{sr}^{y}\tilde{w}_{pr}+
(3-\nu)\sum_{q=1}^{N_{x}+4}\sum_{r=1}^{N_{y}+4}\bar{B}_{pq}^{x}
 \bar{B}_{sr}^{y}\tilde{w}_{qr}\Bigg]
 \\ \,\,\,\ &\hspace{5cm}(p=1,N_{x}; \,\,s=2,3,...,N_{y}-1)   \tag{50}
\end{align*} 
 \begin{align*}   \label{Plate_BMy_dqm}       
M_{y}=&-D\Bigg(\sum_{r=1}^{N_{y}+4}\bar{B}_{sr}^{y}\tilde{w}_{pr}+
 \nu\sum_{r=1}^{N_{x}+4}\bar{B}_{pr}^{x}\tilde{w}_{rs}\Bigg) +\\
 & g^{2}D\Bigg[\sum_{r=1}^{N_{y}+4}\bar{D}_{sr}^{y}\tilde{w}_{pr}+
 \nu\sum_{r=1}^{N_{x}+4}\bar{D}_{pr}^{x}\tilde{w}_{rs}+
(3-\nu)\sum_{q=1}^{N_{x}+4}\sum_{r=1}^{N_{y}+4}\bar{B}_{pq}^{x}
 \bar{B}_{sr}^{y}\tilde{w}_{qr}\Bigg]
 \\ \,\,\,\ &\hspace{5cm}(p=2,3,...,N_{x}-1; \,\,s=1,N_{y})   \tag{51}
\end{align*} 
\noindent Higher order moment:
 \begin{align*}     \label{Plate_HMx_dqm}     
\bar{M}_{x}=&-g^{2}D\Bigg(\sum_{r=1}^{N_{x}+4}\bar{C}_{pr}^{x}\tilde{w}_{rs}+
\nu\sum_{q=1}^{N_{x}+4}\sum_{r=1}^{N_{y}+4}\bar{A}_{pq}^{x}
 \bar{B}_{sr}^{y}\tilde{w}_{qr}\Bigg)
 \\ \,\,\,\ &\hspace{5cm}(p=1,N_{x}; \,\,s=2,3,...,N_{y}-1)   \tag{52}
\end{align*} 
 \begin{align*}     \label{Plate_HMy_dqm}     
\bar{M}_{y}=&-g^{2}D\Bigg(\sum_{r=1}^{N_{y}+4}\bar{C}_{sr}^{y}\tilde{w}_{pr}+
\nu\sum_{q=1}^{N_{x}+4}\sum_{r=1}^{N_{y}+4}\bar{B}_{pq}^{x}
 \bar{A}_{sr}^{y}\tilde{w}_{qr}\Bigg) \\ 
 \,\,\,\ &\hspace{5cm}(p=2,3,...,N_{x}-1; \,\,s=1,N_{y})   \tag{53}
\end{align*}
\noindent Concentrated force at the free corner:
 \begin{align*}    \label{Concentrated_force_Plate_dqm}      
R=&2D(1-\nu)\Bigg(\sum_{q=1}^{N_{x}+4}\sum_{r=1}^{N_{y}+4}\bar{A}_{pq}^{x}
 \bar{A}_{sr}^{y}\tilde{w}_{qr}\Bigg) -\\
 & 2g^{2}D(1-\nu)\Bigg[\sum_{q=1}^{N_{x}+4}\sum_{r=1}^{N_{y}+4}\bar{C}_{pq}^{x}
 \bar{A}_{sr}^{y}\tilde{w}_{qr}+
\sum_{q=1}^{N_{x}+4}\sum_{r=1}^{N_{y}+4}\bar{A}_{pq}^{x}
 \bar{C}_{sr}^{y}\tilde{w}_{qr}\Bigg]
 \\ \,\,\,\ &\hspace{5cm}(p=1,N_{x}; \,\,s=2,3,...,N_{y}-1)   \tag{54}
\end{align*} 

\noindent here $p=1$ and $p=N_{x}$ correspond to the two edges of the plate at $x=0$ and $x=l_{x}$, respectively. Similarly, $s=1$ and $s=N_{y}$ correspond to $y=0$ and $y=l_{y}$ edges.

Once the boundary conditions in Equation (\ref{eq:BC_Cl_Plate_M})-(\ref{eq:BC_NCl_Plate_CForce}), are applied we get a similar system of equations in the matrix form as given by Equation (\ref{eq:Boundary_disp_Beam}). By condensing the boundary dofs, the sysem of equations are reduced to the form as given by Equation (\ref{eq:Domain_disp_Beam}), and the solution leads to the unknown domain displacements of the plate. The boundary displacement are post-processed from Equation (\ref{eq:Boundary_disp_Beam}) and the stress resultants from Equation (\ref{Plate_BMx_dqm})-(\ref{Plate_HMy_dqm}).  \\

In the above we have formulated the differential quadrature plate element based on Lagrange interpolation functions, next, we construct a second version of quadrature plate element based on mixed Lagrange-Hermite interpolation functions.

\subsubsection{Mixed interpolation based differential quadrature element for gradient elastic plates}

The differential quadrature plate element presented here is based on mixed Lagrange-Hermite interpolation, with Lagrangian interpolation is assumed in $x$ direction and Hermite in $y$ direction. The advantage in the mixed interpolation based schemes is the mixed derivative dofs at the free corners of the plate are excluded from the formulation\cite{Wangb}. The deflection for a $N_{x} \times N_{y}$ grid mixed interpolation differential quadrature plate element is assumed as
\begin{align*}    
w(x,y,t)=\sum_{p=1}^{N} \sum_{q=1}^{N+4}L_{p}(x)\Gamma_{q}(y)\tilde{w}_{pq}(t) \tag{55}
\end{align*}

\noindent where $w_{pq}(t)$ is the displacement vector at grid point $(p,q)$, and $\bar{L}_{p}(x)$, $\bar{\Gamma}_{q}(y)$ are the Lagrange and Hermite interpolation functions in $x$ and $y$ directions, respectively. The Lagrange interpolation functions are derived in section \ref{Lagrange_Beam_section}. 
The Hermite interpolation functions for a 1-D N-node differential quadrature element are presented next. The displacement within the 1-D gradient element based on $C^{2}$ continuous Hermite interpolation is assumed as 
\begin{align*} \label{Hermite}     
v(x,y,t)=\sum_{p=1}^{N} \phi_{p}(y)v_{p}+\psi_{1}(y)v_{1}^{'}+\psi_{N}(y)v_{N}^{'}+\varphi_{1}(y)v_{1}^{''}+\varphi_{N}(y)v_{N}^{''}=\sum_{p=1}^{N+4} \Gamma_{p}(y)\bar{v}_{p}\tag{56}
\end{align*}

\noindent where, $\bar{v}$ is the nodal displacement vector, $\phi$, $\psi$ and $\varphi$ are Hermite interpolation functions defined as \cite{Six,Wangnew2}
\begin{align*} \label{Hermite1}   
\varphi_{p}(y)=\frac{1}{2(y_{p}-y_{N-p+1})^{2}}L_{p}(y)(y-y_{p})^{2}(y-y_{N-p+1})^{2}  (p=1, N)\tag{57}
\end{align*}
\begin{align*}    \label{Hermite2}
\psi_{p}(y)=\frac{1}{(y_{p}-y_{N-p+1})^{2}}L_{p}(y)(y-y_{p})(y-y_{N-p+1})^{2}\\
-\bigg[2L_{p}^{1}(y
_{p})+\frac{4}{y_{p}-y_{N-p+1}}\bigg]\varphi
_{p}(x)  \,\,\, (p=1, N)\tag{58}
\end{align*}
\begin{align*}   \label{Hermite3} 
\phi_{p}(y)=\frac{1}{(y_{p}-y_{N-p+1})^{2}}L_{p}(y)(y-y_{N-p+1})^{2} -\bigg[L_{p}^{1}(y
_{p})+\frac{2}{y_{p}-y_{N-p+1}}\bigg]\psi 
_{p}(y)\\-\bigg[L_{p}^{2}(y
_{p})+\frac{4L_{p}^{1}(y
_{p})}{y_{p}-y_{N-p+1}}+\frac{2}{(y_{p}-y_{N-p+1})^{2}}\bigg]\varphi
_{p}(y)  \,\,\, (p=1, N)\tag{59}
\end{align*}
\begin{align*}    \label{Hermite4}
\phi_{p}(y)=\frac{1}{(y_{p}-y_{1})^{2}(y_{p}-y_{N})^{2}}L_{p}(y)(y-y_{1})^{2}(y-y_{N})^{2}   \,\,\, (p=2,3,...,N-1)\tag{60}
\end{align*}

The $n$th order derivative of $v(y,t)$ with respect to $y$ is obtained from Equation (\ref{Hermite}) as 

\begin{align*}    \label{Hermite5}
v^{n}(x,y,t)=\sum_{p=1}^{N} \phi_{p}^{n}(y)v_{p}+\psi_{1}^{n}(y)v_{1}^{'}+\psi_{N}^{n}(y)v_{N}^{'}+\varphi_{1}^{n}(y)v_{1}^{''}+\varphi_{N}^{n}(y)v_{N}^{''}=\sum_{p=1}^{N+4} \mathlarger\Gamma_{p}^{n}(y)\bar{v}_{p}\tag{61}
\end{align*}

Now expressing the Equation (\ref{eq:EOM_Plate}), at inner grid points using the Lagrange and Hermite interpolation, we get
 \begin{align*}         
&D\Bigg[\sum_{r=1}^{N_{x}+4}\bar{D}_{pr}^{x}\tilde{w}_{rs}+
 2\sum_{q=1}^{N_{x}+4}\sum_{r=1}^{N_{y}+4}\bar{B}_{pq}^{x}
\Gamma_{sr}^{2(y)}\tilde{w}_{qr}+
 \sum_{r=1}^{N_{y}+4}\Gamma_{sr}^{4(y)}\tilde{w}_{pr}\Bigg]-\\
 & g^{2}D\Bigg[\sum_{r=1}^{N_{x}+4}\bar{F}_{pr}^{x}\tilde{w}_{rs}+  
 3\sum_{q=1}^{N_{x}+4}\sum_{r=1}^{N_{y}+4}\bar{D}_{pq}^{x}
 \Gamma_{sr}^{2(y)}\tilde{w}_{qr}+
  3\sum_{q=1}^{N_{x}+4}\sum_{r=1}^{N_{y}+4}\bar{B}_{pq}^{x}
 \Gamma_{sr}^{4(y)}\tilde{w}_{qr}+\\ &\hspace{3cm}
 \sum_{r=1}^{N_{y}+4}\Gamma_{sr}^{6(y)}\tilde{w}_{pr}\Bigg]= \bar{q}_{o}(x_{p},y_{s}) \\ \,\,\,\ &\hspace{5cm}(p=2,3,...,N_{x}-1; \,\,s=2,3,...,N_{y}-1)   \\ \tag{62}
\end{align*}
	 
\noindent The boundary forces in Equations (\ref{eq:BC_Cl_Plate_M})-(\ref{eq:BC_NCl_Plate_CForce}), are written as \\
\noindent Shear force:
 \begin{align*}    \label{Plate_SFx_dqm_Hermite}     
V_{x}=&-D\Bigg(\sum_{r=1}^{N_{x}+4}\bar{C}_{pr}^{x}\tilde{w}_{rs}+
(2-\nu)\sum_{q=1}^{N_{x}+4}\sum_{r=1}^{N_{y}+4}\bar{A}_{pq}^{x}
 \Gamma_{sr}^{2(y)}\tilde{w}_{qr}\Bigg) +\\
 & g^{2}D\Bigg[\sum_{r=1}^{N_{x}+4}\bar{E}_{pr}^{x}\tilde{w}_{rs}+
(3-\nu)\sum_{q=1}^{N_{x}+4}\sum_{r=1}^{N_{y}+4}\bar{A}_{pq}^{x}
 \Gamma_{sr}^{4(y)}\tilde{w}_{qr}+
3\sum_{q=1}^{N_{x}+4}\sum_{r=1}^{N_{y}+4}\bar{C}_{pq}^{x}
 \Gamma_{sr}^{2(y)}\tilde{w}_{qr}\Bigg]
 \\ \,\,\,\ &\hspace{5cm}(p=1,N_{x}; \,\,s=2,3,...,N_{y}-1)  \\ \tag{63}
\end{align*} 
 \begin{align*}        \label{Plate_SFy_dqm_Hermite} 
V_{y}=&-D\Bigg(\sum_{r=1}^{N_{y}+4}\Gamma_{sr}^{3(y)}\tilde{w}_{pr}+
(2-\nu)\sum_{q=1}^{N_{x}+4}\sum_{r=1}^{N_{y}+4}\bar{B}_{pq}^{x}
 \Gamma_{sr}^{1(y)}\tilde{w}_{qr}\Bigg) +\\
 & g^{2}D\Bigg[\sum_{r=1}^{N_{y}+4}\Gamma_{sr}^{5(y)}\tilde{w}_{pr}+
(3-\nu)\sum_{q=1}^{N_{x}+4}\sum_{r=1}^{N_{y}+4}\bar{D}_{pq}^{x}
 \Gamma_{sr}^{1(y)}\tilde{w}_{qr}+
3\sum_{q=1}^{N_{x}+4}\sum_{r=1}^{N_{y}+4}\bar{B}_{pq}^{x}
 \Gamma_{sr}^{3(y)}\tilde{w}_{qr}\Bigg]
 \\ \,\,\,\ &\hspace{5cm}(p=2,3,...,N_{x}-1; \,\,s=1,N_{y})  \\  \tag{64}
\end{align*} 

\noindent Bending moment:
 \begin{align*}     \label{Plate_BMx_dqm_Hermite}    
M_{x}=&-D\Bigg(\sum_{r=1}^{N_{x}+4}\bar{B}_{pr}^{x}\tilde{w}_{rs}+
\nu \sum_{r=1}^{N_{y}+4}\Gamma_{sr}^{2(y)}\tilde{w}_{pr}\Bigg) +\\
 & g^{2}D\Bigg[\sum_{r=1}^{N_{x}+4}\bar{D}_{pr}^{x}\tilde{w}_{rs}+
\nu\sum_{r=1}^{N_{y}+4}\Gamma_{sr}^{4(y)}\tilde{w}_{pr}+
(3-\nu)\sum_{q=1}^{N_{x}+4}\sum_{r=1}^{N_{y}+4}\bar{B}_{pq}^{x}
 \Gamma_{sr}^{2(y)}\tilde{w}_{qr}\Bigg]
 \\ \,\,\,\ &\hspace{5cm}(p=1,N_{x}; \,\,s=2,3,...,N_{y}-1)   \tag{65}
 \end{align*} 
 \begin{align*}  \label{Plate_BMy_dqm_Hermite}       
M_{y}=&-D\Bigg(\sum_{r=1}^{N_{y}+4}\Gamma_{sr}^{2(y)}\tilde{w}_{pr}+
 \nu\sum_{r=1}^{N_{x}+4}\bar{B}_{pr}^{x}\tilde{w}_{rs}\Bigg) +\\
 & g^{2}D\Bigg[\sum_{r=1}^{N_{y}+4}\Gamma_{sr}^{4(y)}\tilde{w}_{pr}+
 \nu\sum_{r=1}^{N_{x}+4}\bar{D}_{pr}^{x}\tilde{w}_{rs}+
(3-\nu)\sum_{q=1}^{N_{x}+4}\sum_{r=1}^{N_{y}+4}\bar{B}_{pq}^{x}
 \Gamma_{sr}^{2(y)}\tilde{w}_{qr}\Bigg]
 \\ \,\,\,\ &\hspace{5cm}(p=2,3,...,N_{x}-1; \,\,s=1,N_{y})  \\  \tag{66}
\end{align*} 
\noindent Higher order moment:
 \begin{align*}      \label{Plate_HMx_dqm_Hermite}   
\bar{M}_{x}=&-g^{2}D\Bigg(\sum_{r=1}^{N_{x}+4}\bar{C}_{pr}^{x}\tilde{w}_{rs}+
\nu\sum_{q=1}^{N_{x}+4}\sum_{r=1}^{N_{y}+4}\bar{A}_{pq}^{x}
 \Gamma_{sr}^{2(y)}\tilde{w}_{qr}\Bigg)
 \\ \,\,\,\ &\hspace{5cm}(p=1,N_{x}; \,\,s=2,3,...,N_{y}-1)   \tag{67}
\end{align*} 
 \begin{align*}         \label{Plate_MMy_dqm_Hermite}
\bar{M}_{y}=&-g^{2}D\Bigg(\sum_{r=1}^{N_{y}+4}\Gamma_{sr}^{3(y)}\tilde{w}_{pr}+
\nu\sum_{q=1}^{N_{x}+4}\sum_{r=1}^{N_{y}+4}\bar{B}_{pq}^{x}
 \Gamma_{sr}^{1(y)}\tilde{w}_{qr}\Bigg) \\ 
 \,\,\,\ &\hspace{5cm}(p=2,3,...,N_{x}-1; \,\,s=1,N_{y})  \\  \tag{68}
\end{align*} 
\noindent Concentrated force at the free corner:
 \begin{align*}    \label{Concentrated_force_Plate_dqm}      
R=&2D(1-\nu)\Bigg(\sum_{q=1}^{N_{x}+4}\sum_{r=1}^{N_{y}+4}\bar{A}_{pq}^{x}
\Gamma_{sr}^{1(y)}\tilde{w}_{qr}\Bigg) -\\
 & 2g^{2}D(1-\nu)\Bigg[\sum_{q=1}^{N_{x}+4}\sum_{r=1}^{N_{y}+4}\bar{C}_{pq}^{x}
\Gamma_{sr}^{1(y)}\tilde{w}_{qr}+
\sum_{q=1}^{N_{x}+4}\sum_{r=1}^{N_{y}+4}\bar{A}_{pq}^{x}
\Gamma_{sr}^{3(y)}\tilde{w}_{qr}\Bigg]
 \\ \,\,\,\ &\hspace{5cm}(p=1,N_{x}; \,\,s=2,3,...,N_{y}-1)   \tag{69}
\end{align*} 

\noindent here $p=1$ and $p=N_{x}$ correspond to the two edges of the plate at $x=0$ and $x=l_{x}$, respectively. Similarly, $s=1$ and $s=N_{y}$ correspond to $y=0$ and $y=l_{y}$ edges.

Once the boundary conditions are applied we get a similar system of equations in the matrix form as given in Equation (\ref{eq:Boundary_disp_Beam}), and the solution leads to the unknown displacements of the plate. The stress resultants are obtained by post-processing the Equations (\ref{Plate_BMx_dqm_Hermite})-(\ref{Concentrated_force_Plate_dqm}). 

\section{Numerical Results and Discussion}
 The efficiency of the proposed differential quadrature beam and plate elements is demonstrated for static analysis. First, the performance of the beam element is verified, followed by the plate element. The results reported herein are generated using a single element for different boundary and loading conditions. The classical (deflection, slope and bending moment) and the non-classical (curvature and higher order moment) quantities related to gradient Euler-Bernoulli beam and Kirchhoff plate are compared with the literature results for four values of length scale parameter, $g=0.00001, 0.05, 0.1$, and $0.5$. For ease of comparison, the proposed (strain gradient) differential quadrature beam element based on Lagrange interpolation is designated as SgDQE-L, the plate element based on Lagrange interpolation in $x$ and $y$ directions as SgDQE-LL and the element based on mixed interpolation with Lagrange function in $x$ direction and Hermite function in $y$ direction as SgDQE-LH. 
 
\subsection{Differential quadrature element for gradient elastic Euler-Bernoulli beam}

The  classical and non-classical boundary conditions used in this study for different end supports are listed in the section \ref{Section_Sg_beam}. The non-classical boundary conditions employed for simply supported gradient beam is $w^{''}=0$ at $x=(0,L)$, the equations related to curvature degrees of freedom are eliminated. For the cantilever beam, results are compared for two different choice of non-classical boundary conditions. In the first choice, the non-classical boundary conditions used are $w^{''}=0$ at $x=0$ and $\bar{M}=0$ at $x=L$. The equation related to curvature degrees of freedom at $x=0$ is eliminated and the equation related to higher order moment at $x=L$ is retained. For the next choice of non-classical boundary conditions we assume, $\bar{M}=0$ at $x=0$ and $w^{''}=0$ at $x=L$. Similarly, for clamped and propped cantilever beam the non-classical boundary conditions remains the same, $w^{''}=0$ at $x=(0,L)$. 

In what follows, the ability of the beam element is assessed through convergence study and numerical comparisons for various examples. In the first part, beams subjected to uniformly distributed load (udl) are considered for different support conditions, later, a simply supported and cantilever beam with concentrated load are examined. The numerical data used for the analysis of beams is as follows: Length $L=1$, Young's modulus $E=3 \times 10^{6}$, Poission's ratio $\nu=0.3$ and load $q_{b}=1$. 

\subsubsection{Static analysis of gradient elastic beams under uniformly distributed load}

The results reported here for beams with udl are nondimensional as,  deflection : ${w}_{b}=100EIw/q_{b}L^{4}$, bending moment (BM): $B^{b}_{m}=M/q_{b}L^{2}$, curvature :${w}_{b}^{''}=w^{''}L$ and higher moment : $H^{b}_{m}=\bar{M}/q_{b}L^{3}$. Four support conditions for the beam are considered in this study, simply supported, clamped, cantilever and propped cantilever. In Appendix-I, the procedure used to obtain the exact solutions for different boundary conditions for a gradient elastic Euler-Bernoulli beam under udl is explained. These exact solutions are used to compare the results obtained using SgDQE-L beam element.  

\begin{figure}[!htp]
\includegraphics[width=1.0\textwidth]{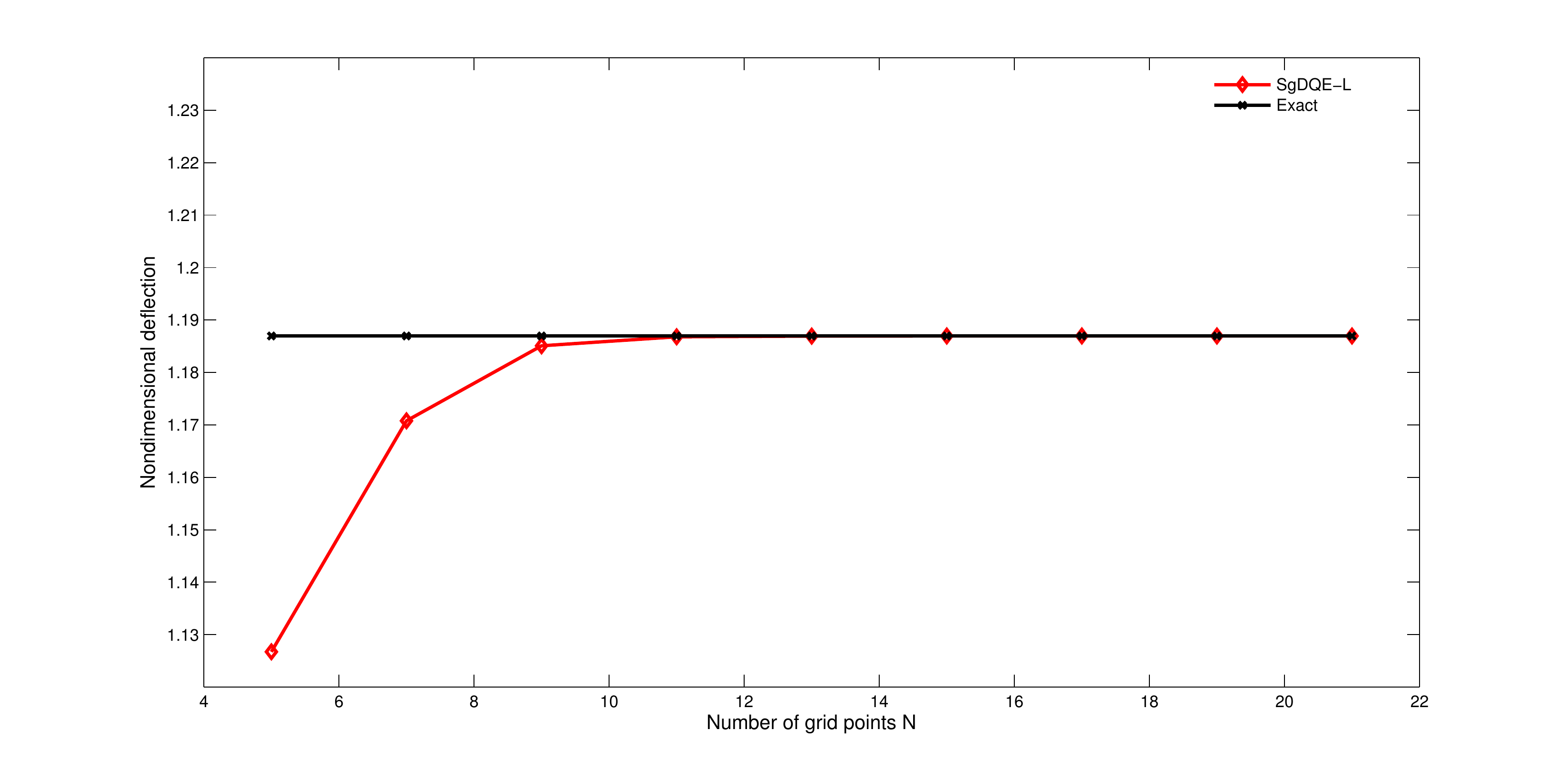}
\centering
\caption{Convergence of maximum nondimensional deflection for a simply supported beam under a udl (g/L=0.1).}
\label{fig:Conv_SS_udl}
\end{figure}

\begin{figure}[!htp]
\includegraphics[width=1.0\textwidth]{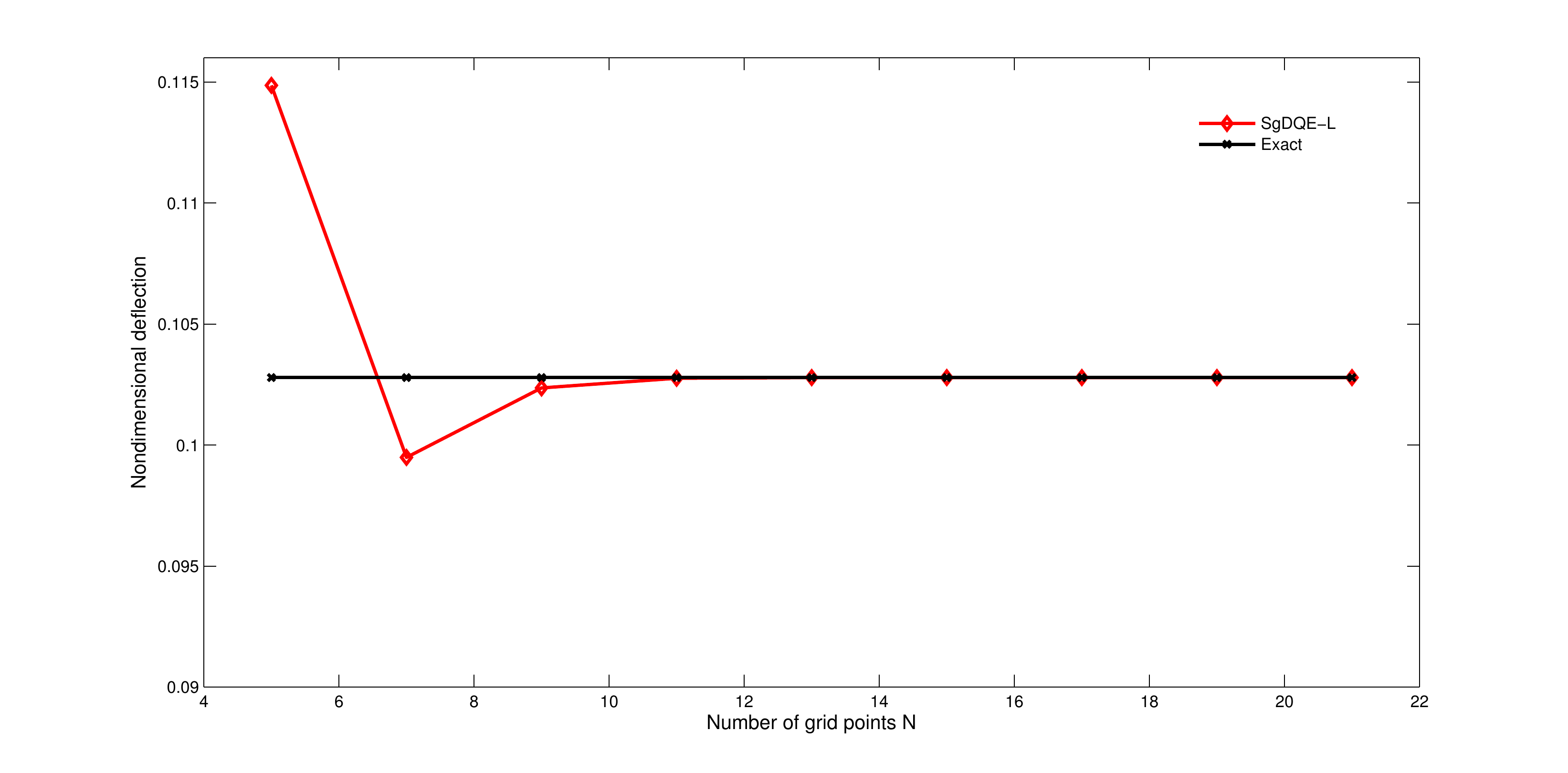}
\centering
\caption{Convergence of maximum nondimensional deflection for a clamped beam under udl (g/L=0.1).}
\label{fig:Conv_Clamped_udl}
\end{figure}

In Figure \ref{fig:Conv_SS_udl}, the convergence of maximum nondimensional deflection obtained using SgDQE-L element for a simply supported gradient beam subjected to udl is shown. The results are compared with exact solutions for $g/L=0.1$. It can be noticed that the convergence is faster for SgDQE-L element, with deflection approaching to exact value with 11 grid points. Similar trend is noticed in the the Figure \ref{fig:Conv_Clamped_udl}, for a clamped beam. Hence, from the above findings, it can be inferred that the accurate solutions can be obtained using single SgDQE-L element with fewer number of nodes.

In the following tables classical and non-classical quantities are compared for different boundary conditions and $g/L$ values. The number of grid points employed to generate the tabulated results is $N=11$. In Table \ref{Behav_at_Point_SS_udl}, the classical and non-classical quantities are given for a simply supported beam and compared with exact solutions for various $g/L$. The deflection and curvature are computed at center of the beam $x=L/2$, the slope and higher order moment at $x=0$. Hence, the classical and non-classical quantities obtained using SgDQE-L are highly accurate and this consistency is maintained for all the $g/L$ values.

\setlength{\extrarowheight}{.5em}
\begin{table}[H]
    \centering
    \caption{Comparison of deflection, slope, curvature and higher order moment for a simply supported beam under a udl.}
      
 \begin{tabular}{p{3.65em}p{4.5em}c@{\hskip 0.3in}c@{\hskip 0.3in}c@{\hskip 0.3in}c@{\hskip 0.3in}}
     \\ \hline
    $ \text{Response}$ & $g/l_{x}$      & 0.00001	         & 0.05           & 0.1     & 0.5  \\ \hline 
${w}_{b\,(x=L/2)}$ 	    & SgDQE-L         &  1.3021  & 1.2702  & 1.1868  & 0.3767   \\
  & Exact   & 1.3021      &  1.2715 & 1.1869 & 0.3767  
       \\ \hline

$w^{'}_{b\,(x=0)}$ 	    & SgDQE-L         &  0.1667  & 0.1620  & 0.1506  & 0.0476 \\ 
$\times 10^{-5}$   & Exact   & 0.1667      &  0.1622 & 0.1507 & 0.0475    \\   \hline

${w}^{''}_{b\,(x=L/2)}$ 	    & SgDQE-L         &  0.5000  & 0.4896  & 0.4605  & 0.1480   \\ 

$\times 10^{-5}$   & Exact   & 0.4999 & 0.4900  &  0.4605 & 0.1480    \\   \hline

$H^{b}_{m\,(x=0)}$  	    & SgDQE-L         &  0.0000  & 1.1188  & 3.9986  & 29.8007   \\
 $\times 10^{-3}$     & Exact   & 0.0000      &  1.1250 & 4.0000 & 29.8007	 \\ \hline 

        \end{tabular}
    \label{Behav_at_Point_SS_udl}
\end{table}
\setlength{\extrarowheight}{.5em}
\begin{table}[!htp]
    \centering
    \caption{Comparison of deflection, slope, curvature, bending moment and higher order moment for a clamped beam under a udl.}
      
 \begin{tabular}{p{4.5em}p{4.5em}c@{\hskip 0.3in}c@{\hskip 0.3in}c@{\hskip 0.3in}c@{\hskip 0.3in}}
     \\ \hline
   $ \text{Response}$ & $g/l_{x}$      & 0.00001	         & 0.05           & 0.1     & 0.5  \\ \hline 
             
${w}_{b\,(x=L/2)}$ 	    & SgDQE-L        &0.2604  & 0.1675  & 0.1028   & 0.0083   \\
      & Exact   & 0.2604 &  0.1678 &  0.1028 &  0.0083  \\  \hline 
      
$w^{'}_{b\,(x=0.2061L)}$    & SgDQE-L         &  3.2059  & 2.1478  & 1.2859  & 0.1003   \\ 
 $\times 10^{-5}$      & Exact   & 3.1759      &  2.0642 & 1.2113 & 0.0931  \\  \hline  
 
${w}^{''}_{b\,(x=L/2)}$  	 & SgDQE-L         &  16.6667  & 12.9666  & 8.8951  & 0.7918  \\ 
 $\times 10^{-5}$ 	      & Exact   & 16.6667      &  12.9663 & 8.8957 & 0.7919  \\  \hline 

	$B^{b}_{m(x=0)}$     & SgDQE-L         &  83.3333  & 90.0131  & 94.0388  &    99.5440    \\
   $\times 10^{-3}$       & Exact   &  83.3324  & 90.0926 & 94.1643  &    99.5440  \\    \hline

	$H^{b}_{m(x=0)}$  		    & SgDQE-L         &  0.0000  & 3.4210  & 5.4176  & 8.1053    \\
$\times 10^{-3}$     & Exact   & 0.0000      &  3.3796 & 5.4156 & 8.1054  \\     \hline 

        \end{tabular}
    \label{Behav_at_Point_Clamped_udl}
\end{table}

The Tables \ref{Behav_at_Point_Clamped_udl} contains the results for a clamped beam. For the clamped beam the deflection and curvature are obtained at $x=L/2$, were as the bending moment and higher order moment is computed at $x=0$ and the slope is evaluated at $x=0.2061L$. From the above tabulated results it can be concluded that the solutions obtained using SgDQE-L element with 11 grid points are in excellent agreement with the exact solutions for all the boundary conditions and $g/L$ values considered. 

\begin{figure}[H]
\includegraphics[width=1.0\textwidth]{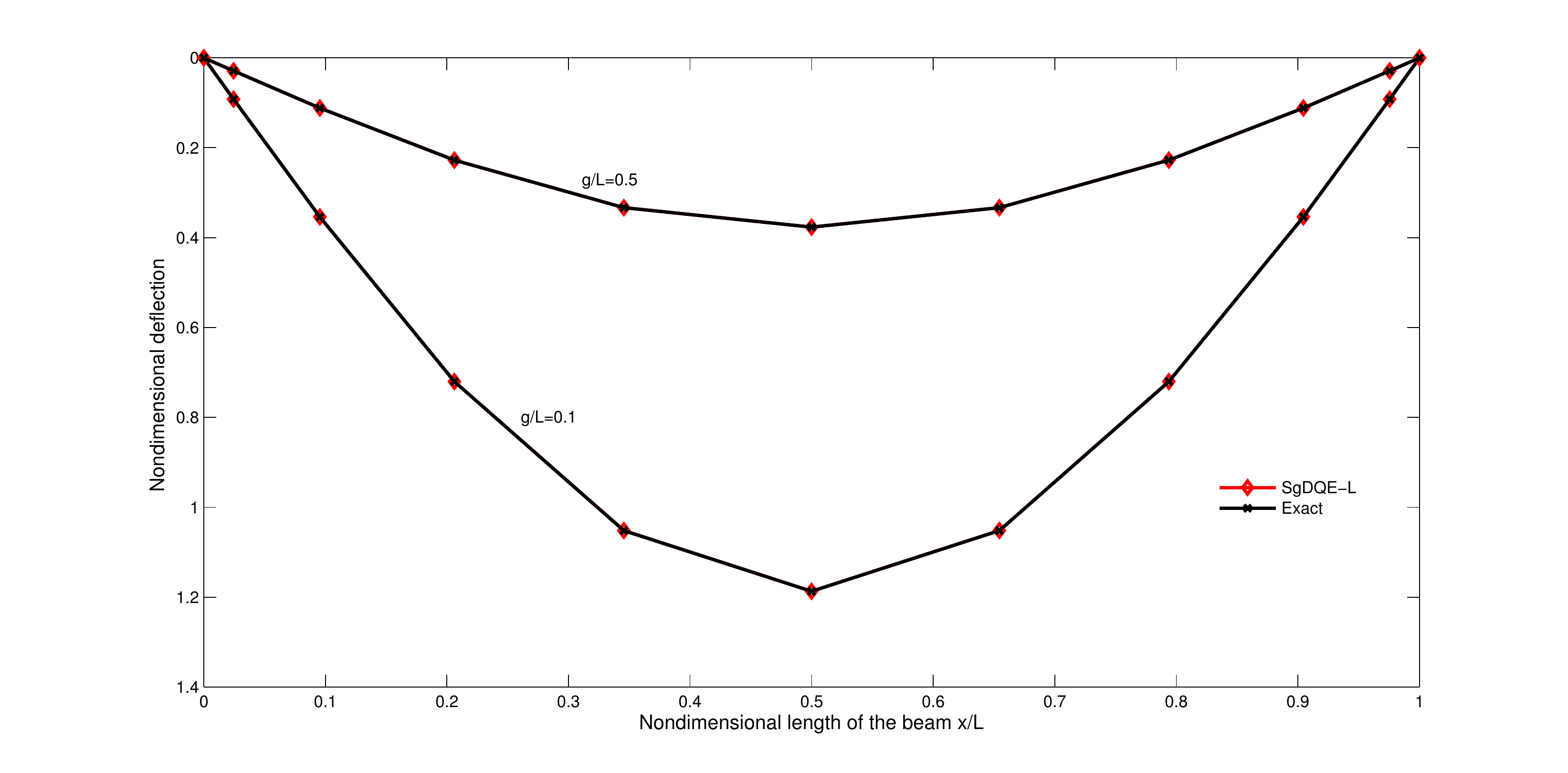}
\centering
\caption{Deflection variation along the length for a simply supported beam under udl.}
\label{fig:Behav_Len_SS_Def_udl}
\end{figure}

\begin{figure}[H]
\includegraphics[width=1.0\textwidth]{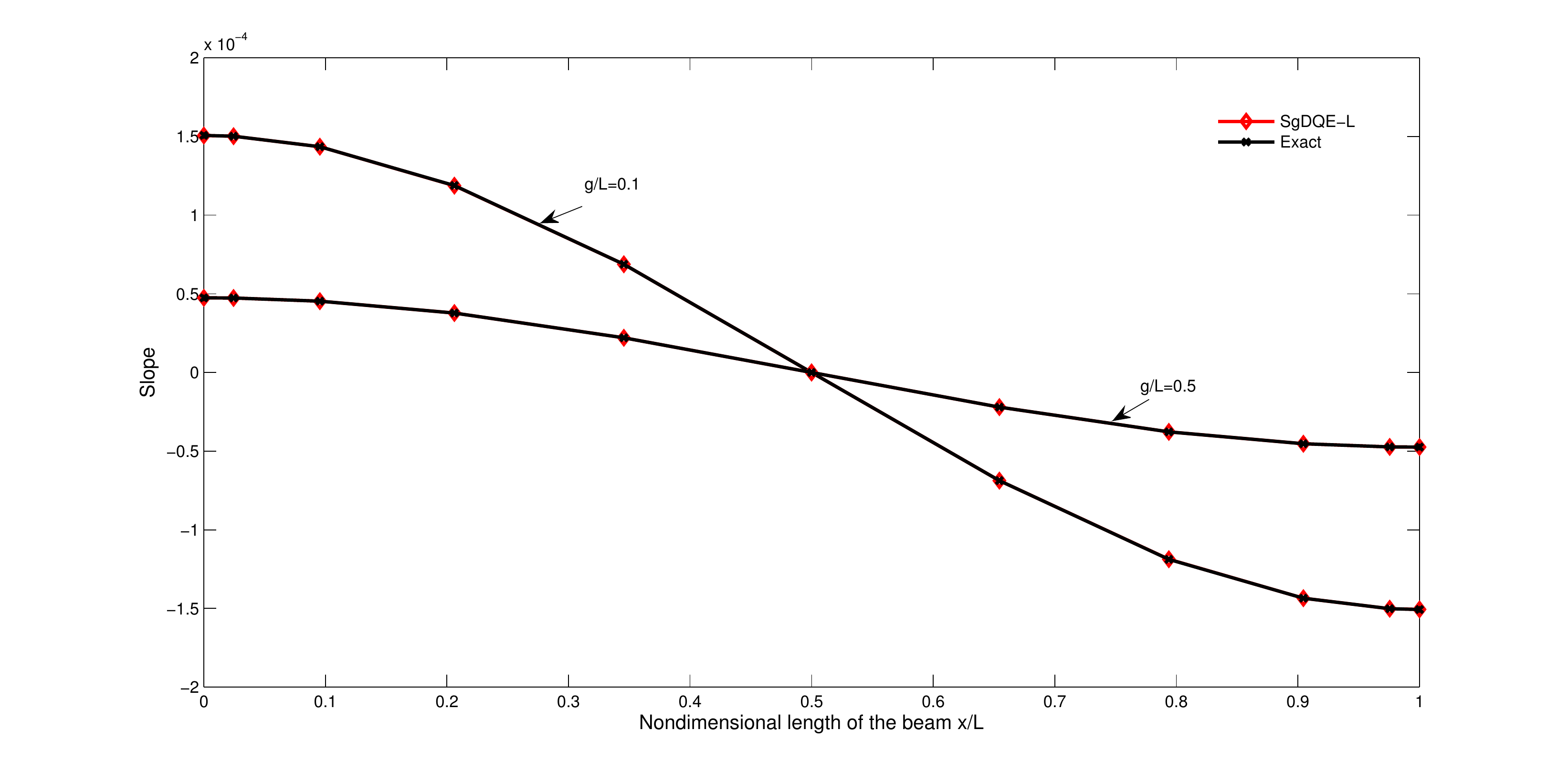}
\centering
\caption{Slope variation along the length for a simply supported beam under udl.}
\label{fig:Behav_Len_SS_Slop_udl}
\end{figure}

\begin{figure}[hp]
\includegraphics[width=1.0\textwidth]{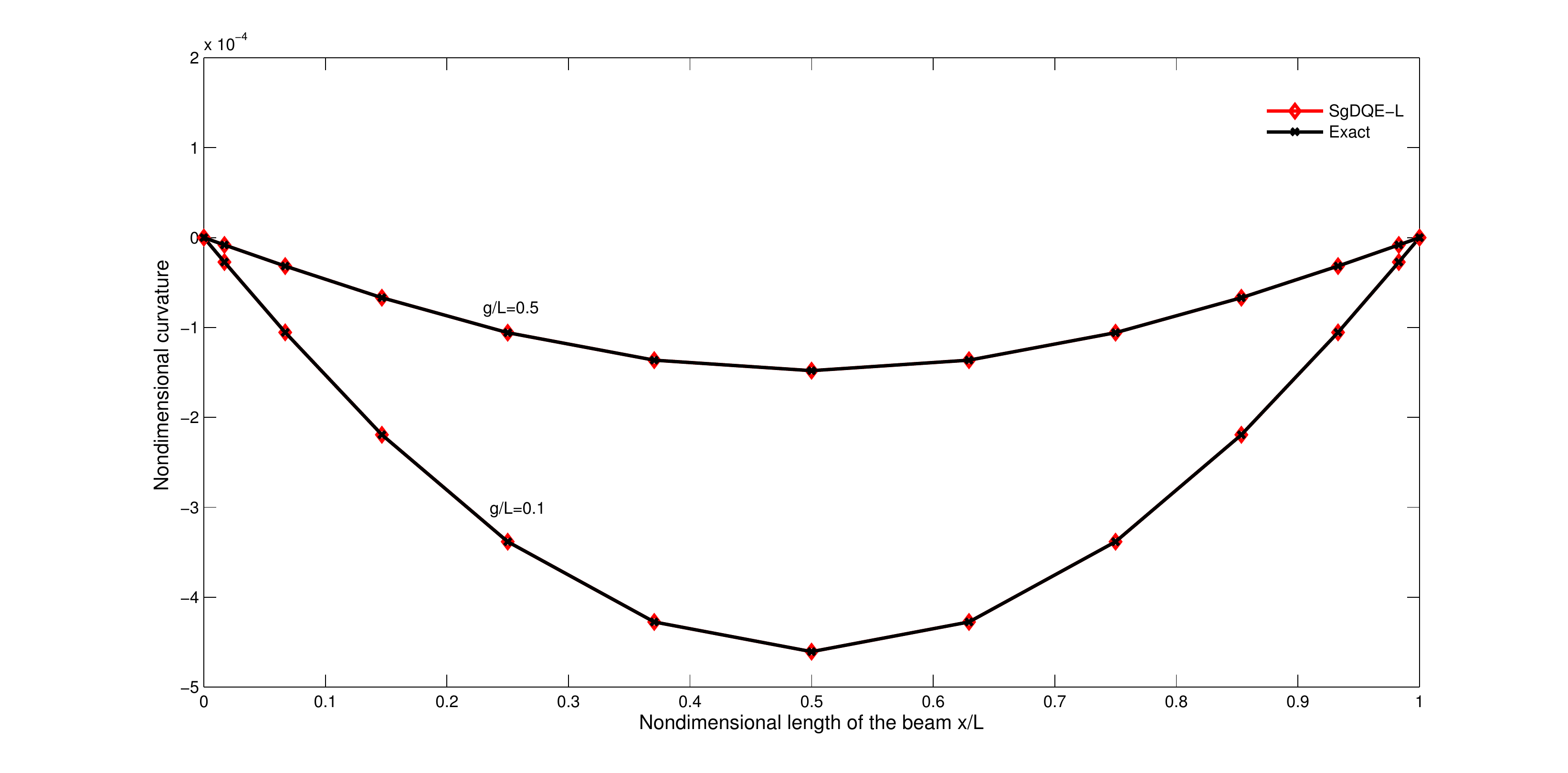}
\centering
\caption{Curvature variation along the length for a simply supported beam under udl.}
\label{fig:Behav_Len_SS_Curv_udl}
\end{figure}

\begin{figure}[H]
\includegraphics[width=1.0\textwidth]{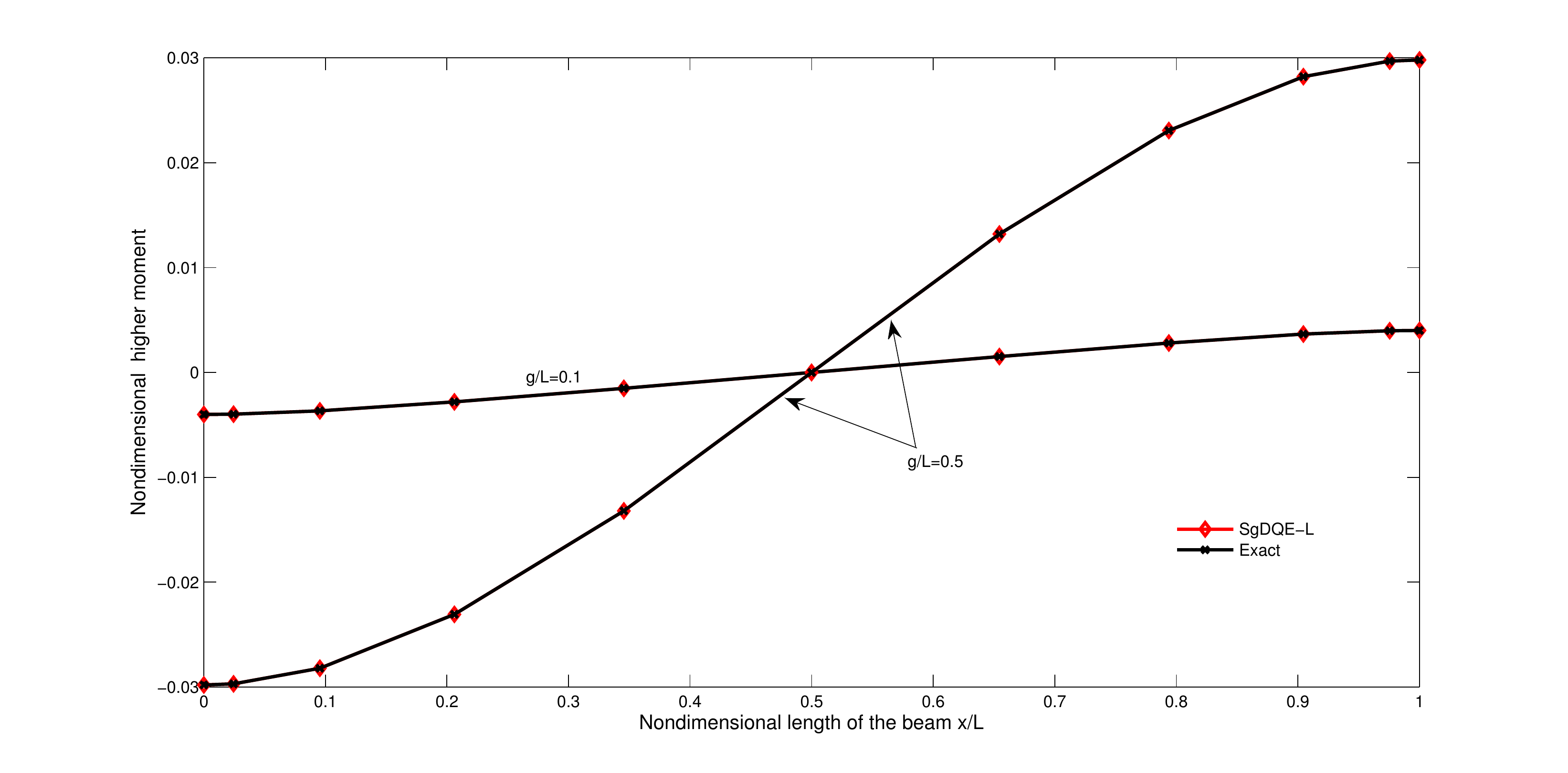}
\centering
\caption{Higher order moment variation along the length for a simply supported beam under udl.}
\label{fig:Behav_Len_SS_Hm_udl}
\end{figure}

In the above results the accuracy of the SgDQE-L element was verified at a particular location of the beam. Next, we demonstrate the accuracy along the length of the beam for $g/L=0.1 \,\,\text{and}\,\, 0.5$. The results are obtained using 11 grid points. Figures \ref{fig:Behav_Len_SS_Def_udl}-\ref{fig:Behav_Len_SS_Hm_udl}, illustrate the variation of deflection, slope, curvature and higher order moment, respectively, along the length for a simply supported beam. The results obtained using SgDQE-L element show perfect fit with the exact solution for both $g/L=0.1$ and $g/L=0.5$. 

From the above observations it can be stated that SgDQE-L element can be efficiently applied to study the static behaviour of gradient elastic Euler-Bernoulli beam for any choice of intrinsic length and boundary condition.

\subsubsection{Static analysis of gradient elastic beams under point load }

\begin{figure}[H]
\includegraphics[width=1.0\textwidth]{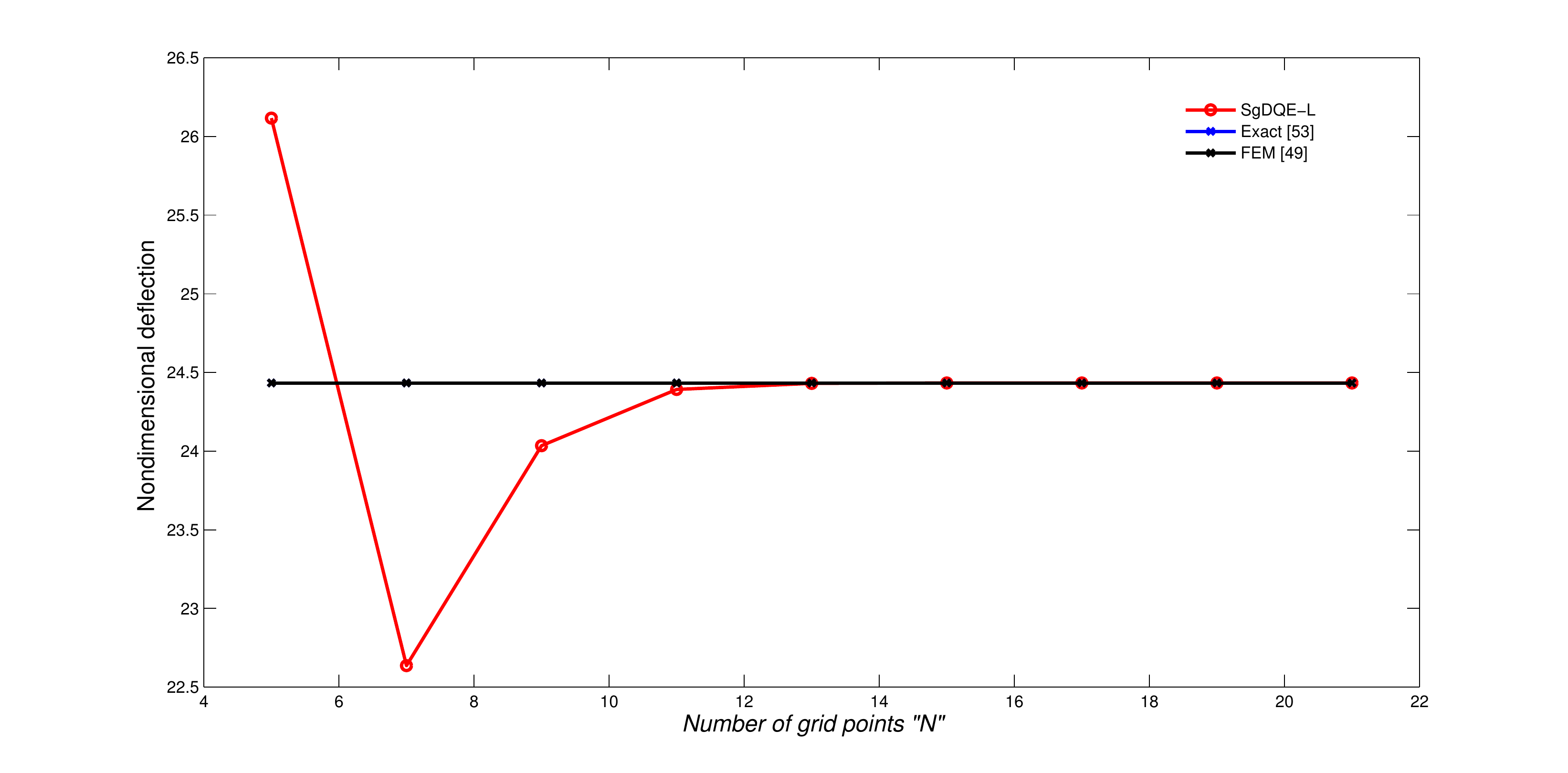}
\centering
\caption{Deflection convergence for a cantilever beam under a tip point load (g/L=0.1).}
\label{fig:Conv_Cant_Pt_load}
\end{figure}

To establish the capability of the SgDQE-L element for beams under concentrated load, two examples are considered, a cantilever beam with tip load and a simply supported beam with mid point load. The results reported here for beams with point load are nondimensional as,  deflection : ${w}_{b}=100EIw/q_{b}L^{3}$, bending moment : $B^{b}_{m}=M/q_{b}L$, curvature : ${w}_{b}^{''}=w^{''}L$ and higher order moment : $H^{b}_{m}=\bar{M}/q_{b}L^{2}$. In the Figure \ref{fig:Conv_Cant_Pt_load}, the convergence of the nondimensional tip deflection for a cantilever beam is shown. The comparison is made with finite element results \cite{Pegios} and exact solutions\cite{Kong}. The SgDQE-L element exhibit excellent and rapid convergence behaviour with 15 grid points. In Table \ref{Behav_Cant_Pt_load}, nondimensional deflection, slope and curvature at the tip for 15 grid points are given. The SgDQE-L element demonstrates good comparison with literature results for all the quantities. 

\setlength{\extrarowheight}{.5em}
\begin{table}[H]
    \centering
    \caption{Comparison of deflection, slope and curvature for a cantilever beam under a tip point load.}
      
 \begin{tabular}{p{3.85em}p{4.5em}c@{\hskip 0.3in}c@{\hskip 0.3in}c@{\hskip 0.3in}c@{\hskip 0.3in}}
     \\ \hline
   $ \text{Response}$ & $g/l_{x}$      & 0.00001	         & 0.05           & 0.1     & 0.5  \\ \hline 
             
    & SgDQE-L        &33.3337  & 28.5784  & 24.4330   & 8.8922   \\
${w}_{b\,(x=L)}$       & Exact\cite{Kong}   & 	33.3323 & 28.5958 & 24.4332 & 8.8922   \\
  & FEM\cite{Pegios}  & 	33.3323 & 28.5958 & 24.4331 & 8.8922   \\
\hline

   & SgDQE-L         & 1.9994  & 1.8086  & 1.6400  & 0.8061   \\ 		 
${w}^{'}_{b\,(x=L)}$     & Exact\cite{Kong}   & 1.9994      &  1.8066 & 1.6266 & 0.8060  \\  $\times 10^{-3}$ 
    & FEM\cite{Pegios}   & 1.9994      &  1.8099 & 1.6400 & 0.8061  \\  

\hline  
       
	 & SgDQE-L         &  0.0016  & 20.3508  & 40.1058  & 86.4905  \\ 
${w}^{''}_{b\,(x=L)}$	      & Exact\cite{Kong}   & 0.0016  & 19.9999    &  39.9636 & 86.4846  \\  
$\times 10^{-5}$  & FEM\cite{Pegios}   & 0.0016  & 19.9999    &  39.9636 & 86.4846  \\    \hline 

        \end{tabular}
     \label{Behav_Cant_Pt_load}
\end{table}

\begin{figure}[H]
\includegraphics[width=1.0\textwidth]{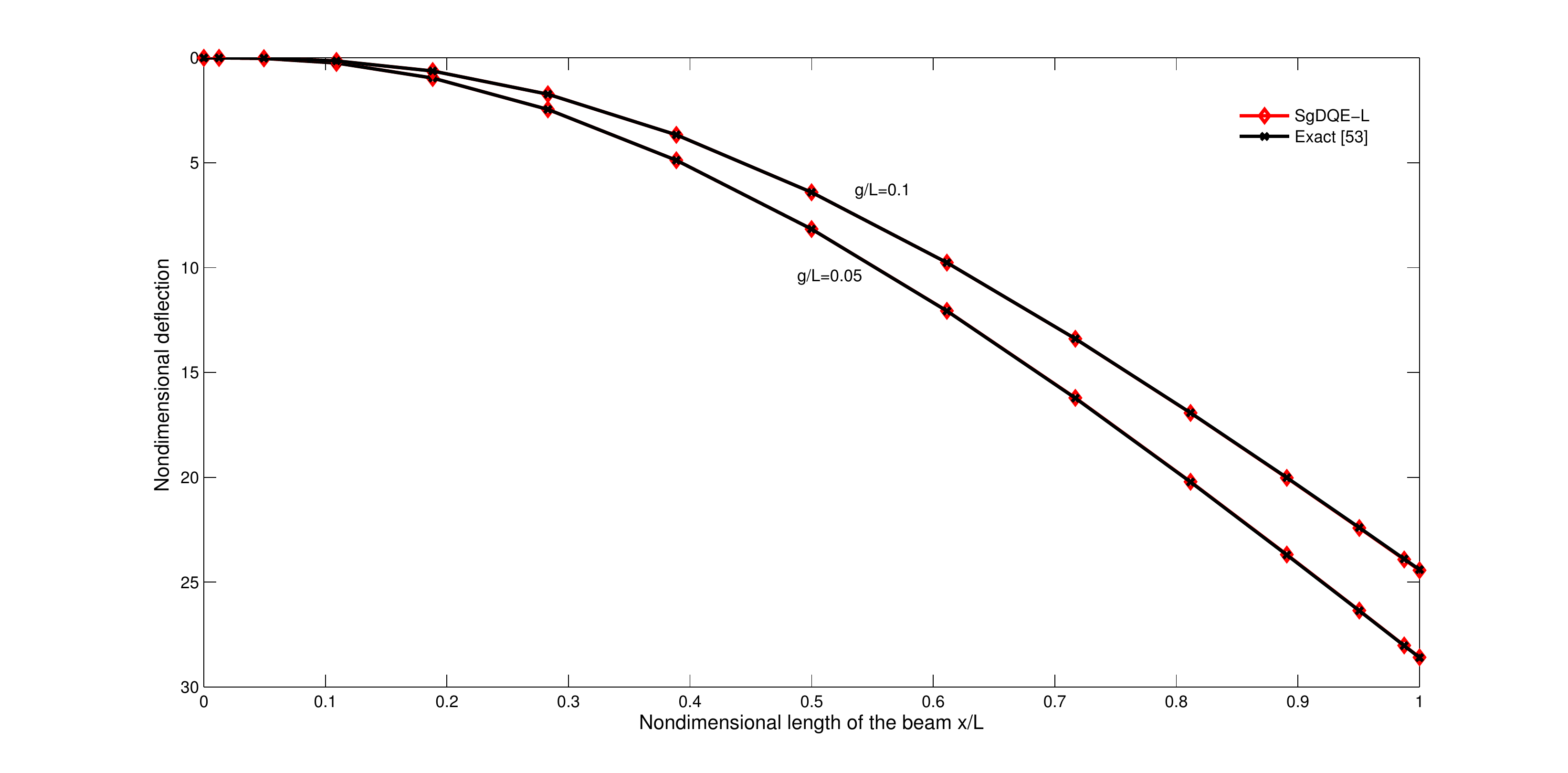}
\centering
\caption{Deflection variation along the length for a cantilever beam under a tip load.}
\label{fig:Behav_Len_Cant_Pt_load}
\end{figure}

To verify the accuracy of the SgDQE-L element along the length of the beam, tip displacement is plotted in Figure \ref{fig:Behav_Len_Cant_Pt_load}, and compared with exact solution \cite{Kong} for $g/L=0.05$ and $0.1$. It is evident from the graph that the SgDQE-L element compares well with the exact solution for both $g/L=0.05$ and $0.1$. 

\begin{figure}[H]
\includegraphics[width=1.0\textwidth]{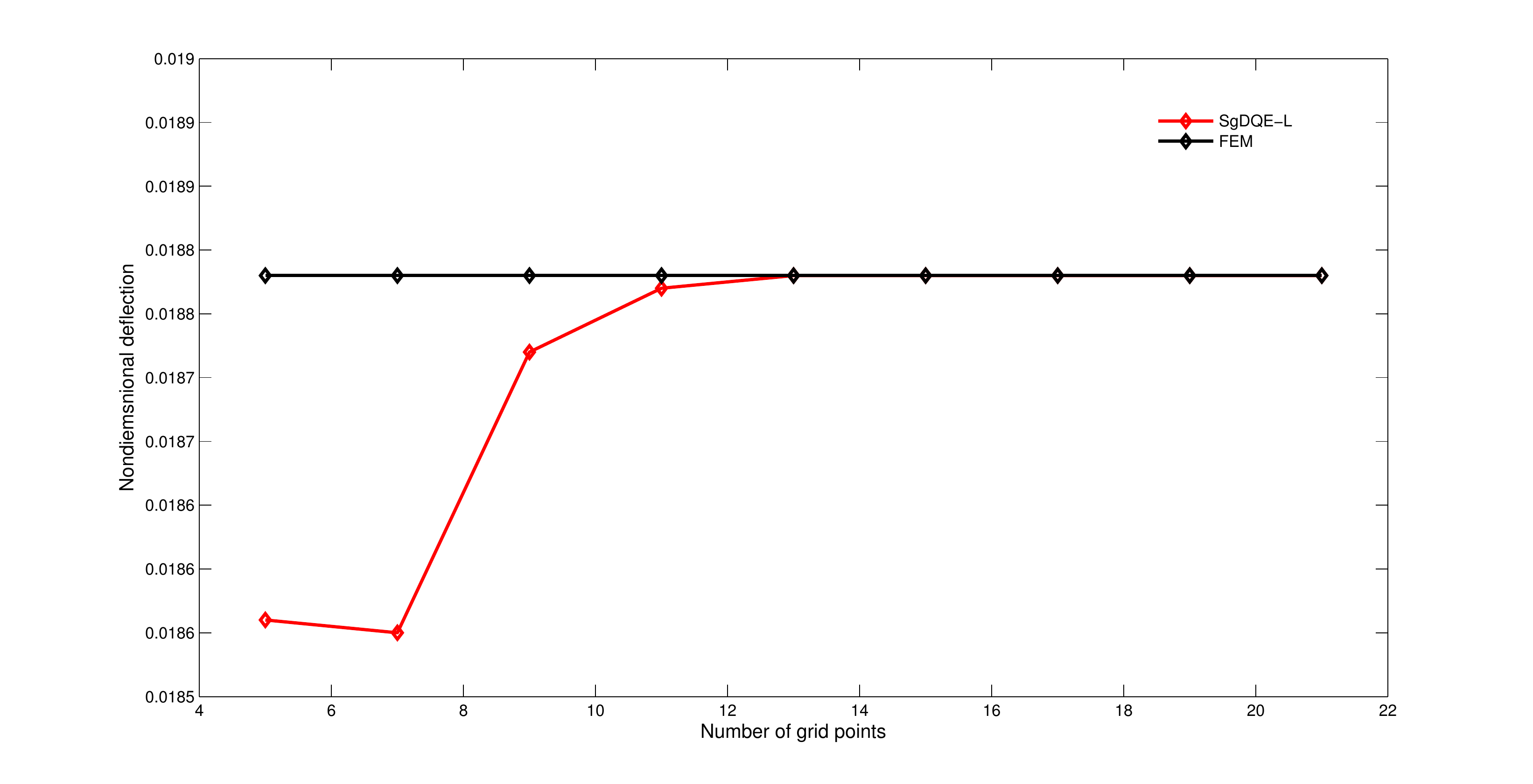}
\centering
\caption{Non-dimensional deflection convergence for a simply supported beam under a mid point load.}
\label{fig:Conv_SS_Pt_load}
\end{figure}

In the Figure \ref{fig:Conv_SS_Pt_load}, the convergence of the nondimensional center deflection for a simply supported beam under mid point load is plotted for $g/L=0.1$. The literature results used to compare the solution are obtained using two finite elements \cite{Pegios}. The Dirac-delta technique is employed here to represent the concentrated load accurately \cite{Eftekari,Wangb} and similar grid is used as in Ref.\cite{Eftekari} to produce the results. The observations indicate the SgDQE-L element display faster convergence with 13 grid points. In Table \ref{Behav_SS_Pt_load}, nondimensional deflection, slope and curvature are tabulated for comparison. The SgDQE-L results agree well with the finite element solutions for all the quantities and $g/L$ values.

Hence, single SgDQE-L element with less number of grid points yield accurate results for beams with concentrated load.

\setlength{\extrarowheight}{.5em}
\begin{table}[H]
    \centering
    \caption{Comparison of deflection, slope and curvature for a simply supported beam under a mid point load.}
      
 \begin{tabular}{p{3.65em}p{4.5em}c@{\hskip 0.3in}c@{\hskip 0.3in}c@{\hskip 0.3in}c@{\hskip 0.3in}}
     \\ \hline
   $ \text{Response}$ & $g/l_{x}$      & 0.00001	         & 0.05           & 0.1     & 0.5  \\ \hline 
             
${w}_{b\,(x=L)}$	  	    & SgDQE-L        & 0.0208  & 0.0267  & 0.01883   & 0.0059   \\
       & Exact   & 0.0208  & 0.0267  & 0.01883   & 0.0059 \\
\hline 

${w}^{'}_{b\,(x=L)}$    & SgDQE-L         & 0.2499  & 0.2450  & 0.2303  & 0.0740   \\ 		 
$\times 10^{-3}$     & Exact   & 0.2500      &  0.2450 & 0.2303 & 0.0740  \\  
\hline  
       
${w}^{''}_{b\,(x=L)}$ 	 & SgDQE-L         &  0.9349  & 0.8854  & 0.7950  & 0.2382  \\ 
$\times 10^{-3}$ 	      & Exact & 0.9999  & 0.9000  & 0.8000    &  0.2384 
 \\
  \hline 

        \end{tabular}
    \label{Behav_SS_Pt_load}
\end{table}

\subsection{Differential quadrature element for a gradient elastic Kirchhoff plate} 

In this section, the performance of the two proposed differential quadrature plate elements SgDQE-LL and SgDQE-LH is validated. Three different example problems are considered here, a rectangular plate with udl, a square plate subjected to central point load and a plate under cylindrical bending. Comparison is made with available literature results for different boundary conditions and $g/l_{x}$ values. For the example problems were gradient plate solutions are not available in the literature, comparison is made with the classical solutions. The boundary conditions used in the examples are described by a notation, for example, a cantilever plate is represented as CFFF, the first and second letter correspond to $x=0$ and $y=0$ edges, similarly, the third and fourth letter correspond to the edges $x=l_{x}$ and $y=l_{y}$, respectively. Further, the letter S, C and F correspond to simply supported, clamped and free edges of the plate. The number of grid points in either direct are assumed to be equal, $N=N_x=N_y$.  The numerical data used for the analysis of plates is: length $l_{x}=1$, width $l_{y}=1$, thickness $h=0.01$, Young's modulus $E=3 \times 10^{6}$, Poission's ratio $\nu=0.3$ and load $\bar{q}_{o}=1$. The deflections reported herein are nondimensional as, $\bar{w}_{p}=100D\bar{w}/\bar{q}_{0}l_{x}^{4}$ for udl and $\bar{w}_{p}=100Dw/\bar{q}_{0}l_{x}^{2}$ for point load. Similarly, the bending moment and higher order moment are nondimensional as, $B^{p}_{m}=M_{x}/\bar{q}_{0}l_{x}^{2}$ and $H^{p}_{m}=\bar{M}_{x}/\bar{q}_{0}l_{x}^{3}$, respectively. The curvature is nondimensional as curvature : ${w}_{p}^{''}=w^{''}l_{x}$

The non-classical boundary conditions employed for SSSS gradient plate are $w_{xx}=0$ at $x=(0,l_{x})$  and  $w_{yy}=0$ at $y=(0,l_{y})$, the equations related to curvature degrees of freedom are eliminated. The size of the resulting non-zero boundary degrees of freedom vector $\Delta_{b}$ after eliminating the equations related zero classical and non-classical boundary degrees of freedom is $4N-8$, which is similar to classical simply supported plate. For a CCCC plate the non-classical boundary conditions used are $w_{xx}=0$ at $x=(0,l_{x})$  and  $w_{yy}=0$ at $y=(0,l_{y})$, as a result, the size of boundary degrees of freedom vector is zero. Similarly, for CFFF cantilever plate, conditions employed are $w_{xx}=0$ at $x=0$, $\bar{M}_{x}=0$ at $x=l_{x}$  and  $\bar{M}_{y}=0$ at $y=(0,l_{y})$, and this leads to the size of $\Delta_{b}$ as $9N-8$.

\subsubsection{Static analysis of gradient Kirchhoff plates under uniformly distributed load}
 
\begin{figure}[H]
\includegraphics[width=1.0\textwidth]{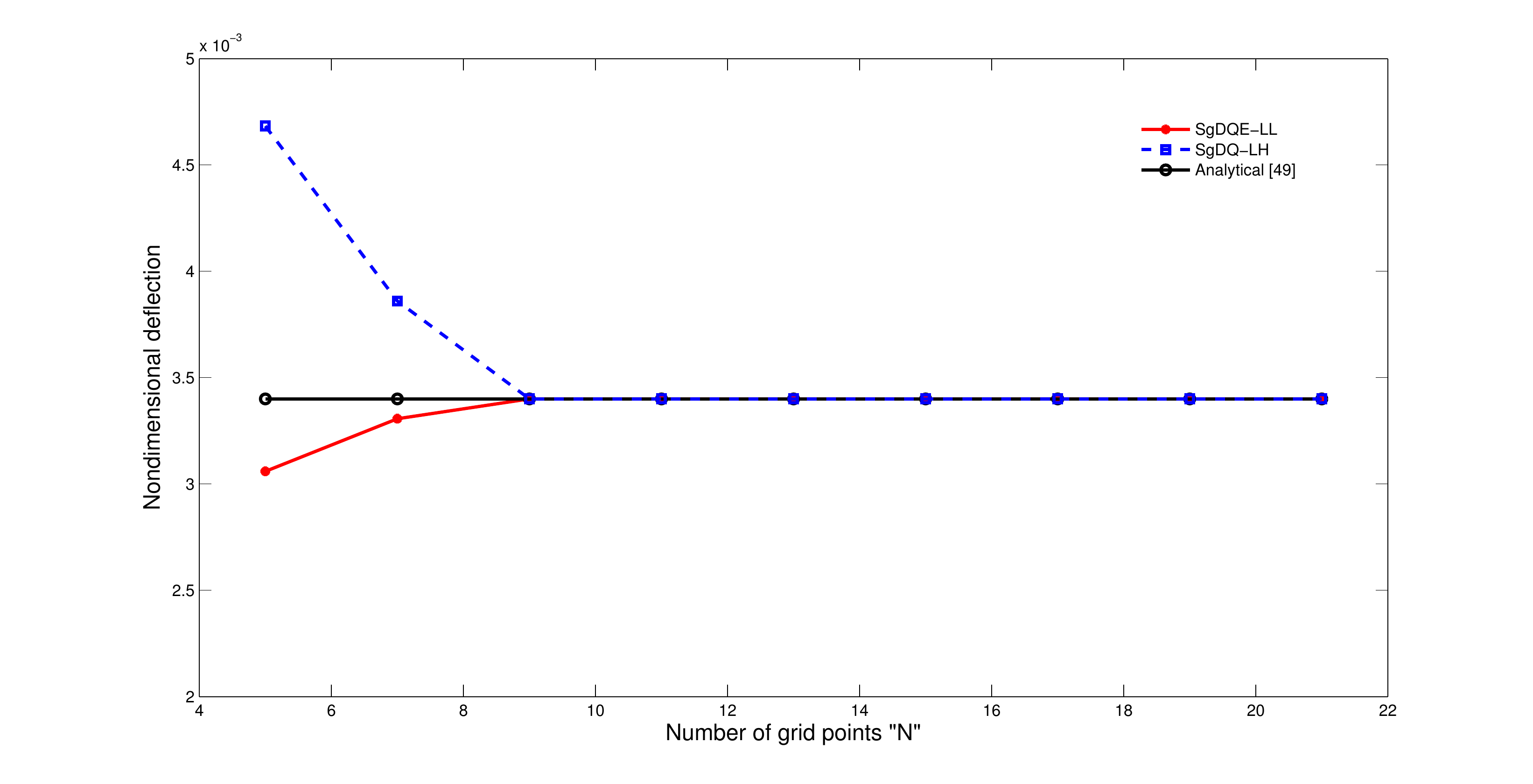}
\centering
\caption{Deflection convergence for a SSSS gradient elastic plate under a udl $(g/l_{x}=0.1)$}
\label{fig:Conv_SSSS_Plate_udl}
\end{figure}

\begin{figure}[H]
\includegraphics[width=1.0\textwidth]{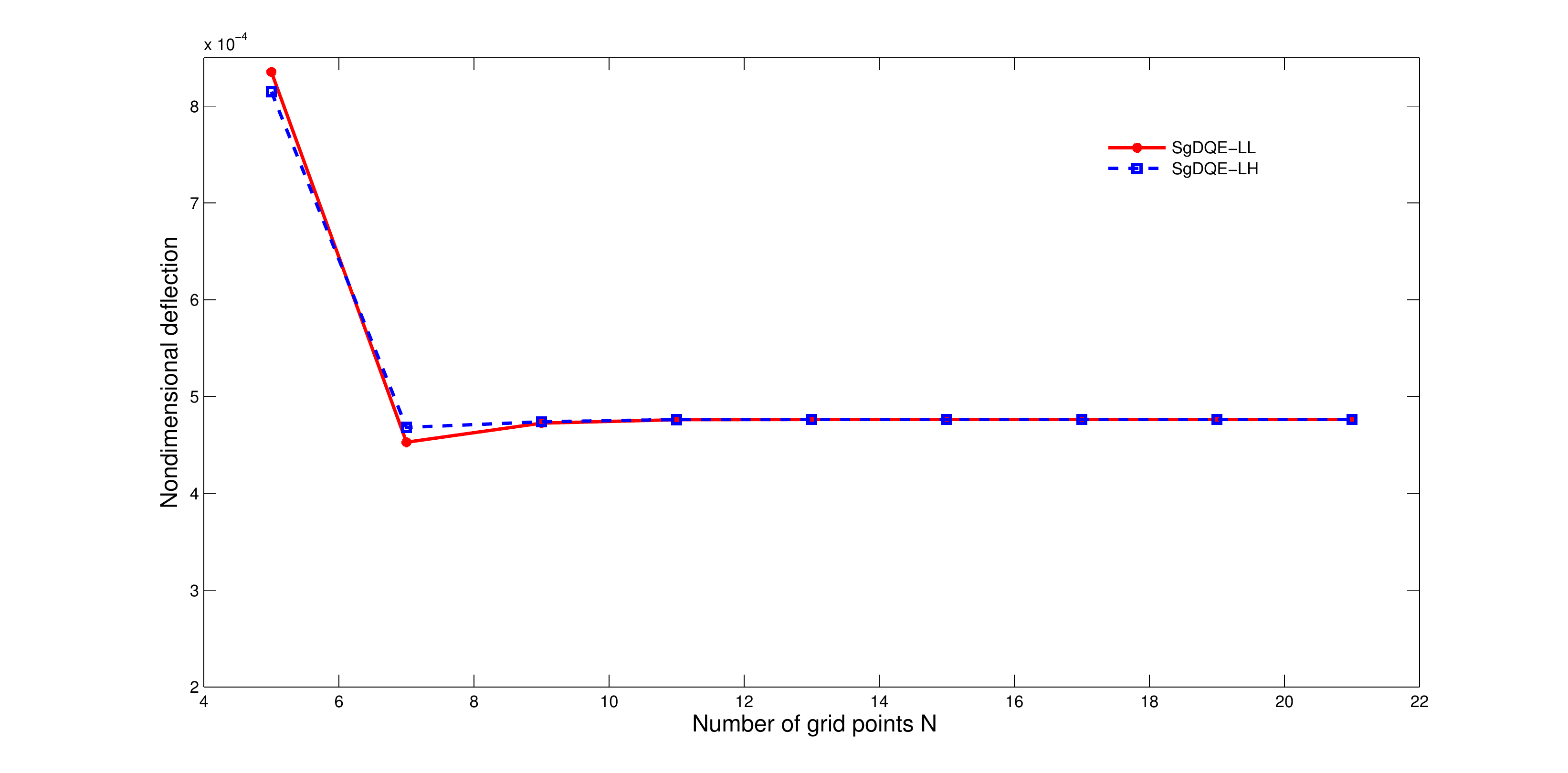}
\centering
\caption{Deflection convergence for a CCCC gradient elastic plate under a udl $(g/l_{x}=0.1)$.}
\label{fig:Conv_CCCC_Plate_udl}
\end{figure}

\begin{figure}[H]
\includegraphics[width=1.0\textwidth]{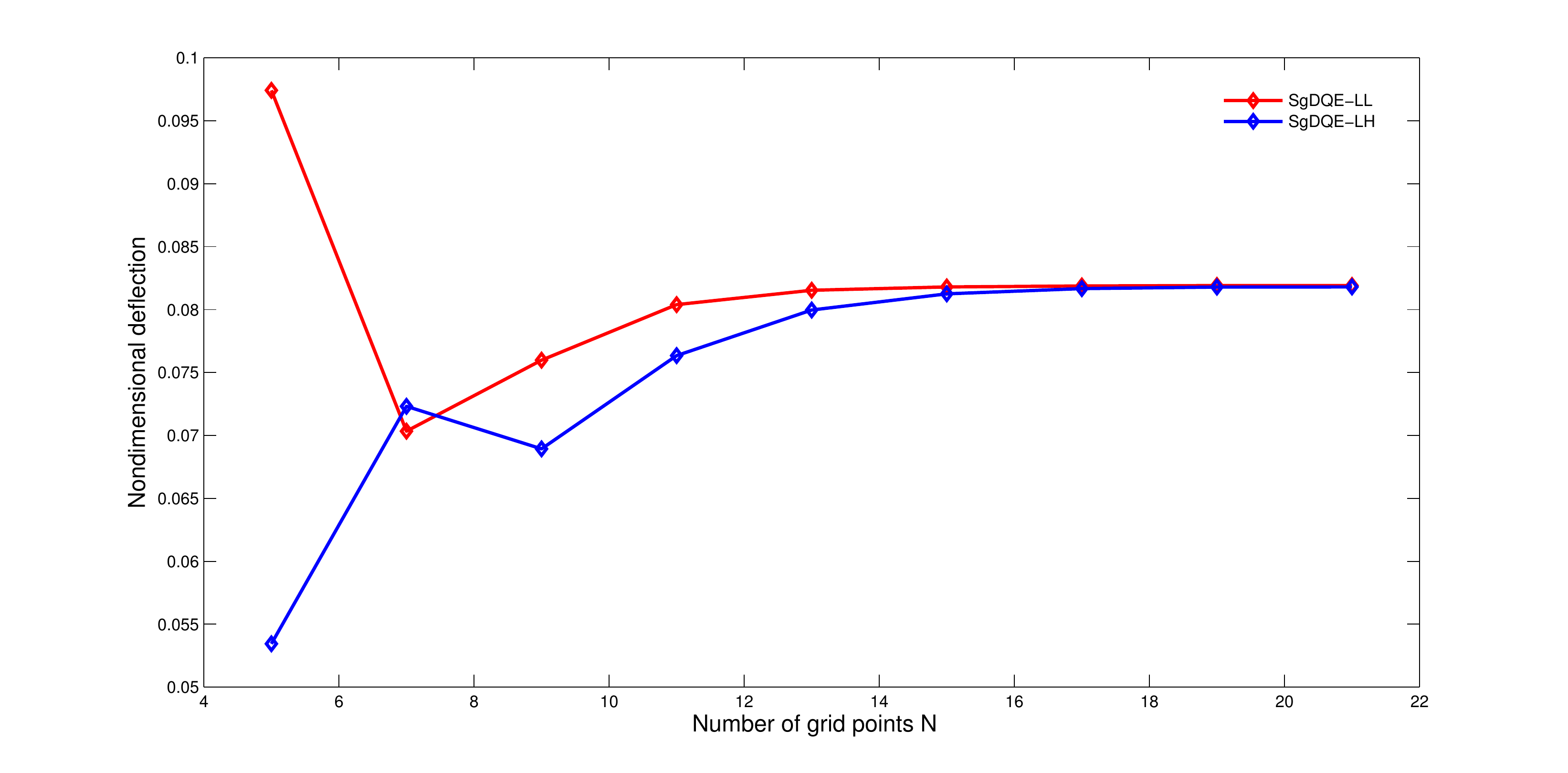}
\centering
\caption{Deflection convergence for a CFFF gradient elastic plate under a udl $(g/l_{x}=0.1)$.}
\label{fig:Conv_FFFC_Plate_udl}
\end{figure}

In this study, a SSSS, CCCC square plate and a rectangular cantilever plate CFFF $(l_{y}=2\,l_{x})$ under udl are analysed. In Figure \ref{fig:Conv_SSSS_Plate_udl}, the convergence of nondimensional deflection for a SSSS plate with $g/l_{x}=0.1$ is plotted for SgDQE-LL and SgDQE-LH elements and compared with analytical solutions\cite{Besko2p}. The analytical solutions are obtained using 100 terms in the series. The SgDQE-LL and SgDQE-LH elements demonstrate faster convergence to exact solution with $N_x=N_y=11$, grid points. In Figures \ref{fig:Conv_CCCC_Plate_udl} and  \ref{fig:Conv_FFFC_Plate_udl}, the convergence behaviour for CCCC and CFFF plates obtained using SgDQE-LL and SgDQE-LH are illustrated. As the analytical solutions for gradient elastic CCCC and CFFF plates are not available in the literature, only the convergence trend for SgDQE-LL and SgDQE-LH elements are shown. Both the elements converge with 11 and 17 grid points for CCCC and CFFF plates respectively. The deflection for SSSS and CCCC plate is computed at centre $(l_{x}/2,\, l_{y}/2)$ .and for CFFF plate at $(l_{x},\,l_{y}/2)$.

In Table \ref{Behav_All_Disp_Plate_udl}, the nondimensional deflection for a SSSS plate computed for different $g/l_{x}$ values is shown. The SgDQE-LL and SgDQE-LH elements demonstrate excellent agreement with the analytical solutions \cite{Besko2p} for all $g/l_{x}$ values. For the CCCC and CFFF plate the classical solution is compared with SgDQE-LL and SgDQE-LH results for lower value of $g/l_{x}=0.00001$. However, for higher values of $g/l_{x}$ only solutions obtained using SgDQE-LL and SgDQE-LH are tabulated and they show good comparison. Similar accuracy is seen in Table \ref{Behav_All_Slope_Plate_udl}-\ref{Behav_All_Hm_Plate_udl}, for slope, curvature and higher order moment, respectively, obtained for different boundary conditions of the plate. All the results presented here for SSSS and CCCC plates are obtained using $N_{x}$=$N_{y}$=15, and for CFFF plate $N_{x}$=$N_{y}$=17 is used. 

In the above, the classical and non-classical quantities are compared with the analytical solutions for SSSS plate and excellent match is verified. For the example problems were analytical solutions are not available, SgDQE-LL and SgDQE-LH elements produce identical results for CCCC and CFFF plates. From the above findings, it can be ascertain that the SgDQE-LL and SgDQE-LH elements with less number of grid points can be efficiently applied to study the static behaviour of gradient plates under udl. Next, we illustrate the applicability of the above elements for plates with concentrate loads.
 
\setlength{\extrarowheight}{.5em}
\begin{table}[H]
    \centering
    \caption{Comparison of nondimensional deflection for a gradient plate under a udl.}
      
 \begin{tabular}{p{3.65em}p{5.5em}c@{\hskip 0.3in}c@{\hskip 0.3in}c@{\hskip 0.3in}c@{\hskip 0.3in}}
     \\ \hline
   $ \text{Support}$ & $g/l_{x}$      & 0.00001	         & 0.05           & 0.1     & 0.5  \\ \hline 

SSSS 	 	    & SgDQE-LL         &  0.4062  &  0.3884   & 0.3423  &  0.0697   \\ 
$w_{p}(\frac{l_{x}}{2},\frac{l_{y}}{2})$  & SgDQE-LH       & 0.4062 &   0.3902 &   0.3481 & 0.0763  \\
        & Ref.\cite{Besko3p}   &   0.4062  & 0.3884 &  0.3423 &  0.0697  \\  \hline

  CCCC   & SgDQE-LL      &     0.1265 &  0.0803 &   0.0476 &   0.0036  \\ 
  $w_{p}(\frac{l_{x}}{2},\frac{l_{y}}{2})$    & SgDQE-LH      &   0.1265 &   0.0803 &   0.0476 &   0.0036  \\ 
& Exact\cite{Wangb} 
         & 0.1265 &  ------ &  ------ &  ------  \\ \hline 
  
  CFFF    & SgDQE-LL      &   12.7770 &   10.3975  & 8.5059 &  2.6490  \\ 
  $w_{p}(l_{x},\frac{l_{y}}{2})$   & SgDQE-LH      &  12.7742  & 10.3434 &  8.4848  & 2.6888  \\ 
& Ref.\cite{Wangold} 
          & 12.87 &  ------ &  ------ &  ------  \\ \hline  
   
           \end{tabular}
    \label{Behav_All_Disp_Plate_udl}
\end{table}

\begin{table}[h]
    \centering
    \caption{Comparison of slope for a gradient plate under a udl.}
      
    \begin{tabular}{p{5.2em}p{5em}p{3.75em}p{3.75em}p{3.75em}p{3.75em}p{3.75em}}
     \\  \hline
support        & $g/l_{x}$      & 0.00001	         & 0.05           & 0.1     & 0.5   \\  \hline  
	SSSS  	    & SgDQE-LL         &  4.9074  &  4.6455   & 4.0407  &  0.8067   \\ 
$w^{'}_{p}(l_{x},\frac{l_{y}}{2})$ & SgDQE-LH       &  4.9075  & 	 4.7177   & 4.0426  &  0.8013    \\ 
       & Ref.\cite{Besko3p}   &   4.7579  & 4.5341 &  3.9735 &  0.8017  \\ \hline 
     
  CCCC & SgDQE-LL      &   1.2424 & 0.6417 &  0.3276 &   0.0218  \\ 
 $w^{'}_{p}(0.188l_{x},\frac{l_{y}}{2})$      & SgDQE-LH      & 1.2424 &  0.6417 &  0.3276 &   0.0218   \\ \hline  
 
CFFF    & SgDQE-LL      &   61.9813 &   53.1975  & 46.2014 &  20.3113 \\ 
   $w^{'}_{p}(l_{x},\frac{l_{y}}{2})$    & SgDQE-LH      &  61.9832 &   52.7951  & 45.9788 &  20.6468 \\  \hline  
           \end{tabular}
    \label{Behav_All_Slope_Plate_udl}
\end{table}

\begin{table}[H]

    \centering
    \caption{Comparison of nondimensional curvature for a gradient plate under a udl.}
  \setlength{\extrarowheight}{.5em}    
    \begin{tabular}{p{3.5em}p{5em}p{3.75em}p{3.75em}p{3.75em}p{3.75em}p{3.75em}}
     \\  \hline
support        & $g/l_{x}$      & 0.00001	         & 0.05           & 0.1     & 0.5   \\  \hline  
	SSSS 	    & SgDQE-LL         &  13.4080  & 12.9528   & 11.6342 &  2.4499   \\ 
$w^{''}_{p}{(\frac{l_{x}}{2},\frac{l_{y}}{2})}$  & SgDQE-LH       &  13.4090  &  12.9896   & 11.7773  &  2.6562   \\ 
    & Ref.\cite{Besko3p}   &   13.4082  & 12.9533 &  11.6342 &  2.4499  \\ \hline 
       
CCCC & SgDQE-LL      &     6.4122 &  4.9539 &   3.3135 &   0.2766  \\ 
 $w^{''}_{p}{(\frac{l_{x}}{2},\frac{l_{y}}{2})}$ & SgDQE-LH      &  6.4123 &  4.9628 &   3.3151 &   0.2766 \\ \hline  
 
  CFFF   & SgDQE-LL      &   0.4740 &   1.2650  & 3.8418 &  18.6090 \\ 
 $w^{''}_{p}{(l_{x},\frac{l_{y}}{2})}$     & SgDQE-LH      &  0.5423 &   1.2650  & 2.8795 &  18.9998  \\  \hline  
           \end{tabular}
    \label{Behav_All_Curv_Plate_udl}
\end{table}

\begin{table}[H]
    \centering
    \caption{Comparison of nondimensional higher order moment for a gradient plate under a udl.}
      
    \begin{tabular}{p{3.5em}p{5em}p{3.75em}p{3.75em}p{3.75em}p{3.75em}p{3.75em}}
     \\  \hline
support        & $g/l_{x}$      & 0.00001	         & 0.05           & 0.1     & 0.5   \\  \hline  
	SSSS 	    & SgDQE-LL         &  0.0000  &  0.5194   & 1.6666  &  7.5376   \\ 
 $H^{p}_{m}(0,\frac{l_{y}}{2})$   & SgDQE-LH       &  0.0000  &  0.5189   & 1.6662  &  7.5367   \\ 
       & Ref. \cite{Besko3p}   &   0.0000  & 0.3999 &  1.4006 &  7.0646  \\  \hline        
       
  CCCC  & SgDQE-LL      &     0.0000 &  2.0802 &   3.3107 &   4.8114  \\ 
 $H^{p}_{m}(0,\frac{l_{y}}{2})$     & SgDQE-LH      &        0.0000 &  2.0781 &   3.3133 &   4.8191   \\\hline  
 
  CFFF   & SgDQE-LL      &   0.0000 &   23.0985  & 41.6790 &  113.1579 \\ 
   $H^{p}_{m}(0,\frac{l_{y}}{2})$   & SgDQE-LH      &  0.0000  & 22.9233 &  41.5215  & 114.2411  \\  \hline  
           \end{tabular}
    \label{Behav_All_Hm_Plate_udl}
\end{table}

\subsubsection{Static analysis of gradient Kirchhoff plate under point load}

To represent the concentrate load accurately the Dirac-delta technique is employed \cite{Wangb, Eftekari}. In Table \ref{Behav_All_Disp_Plate_Ptl}, the nondimensional defection for CFCF, SFSF and SFCF plates subjected to central point load are presented. Simlar grid is used as in Ref.\cite{Wangb} to generate the results. The solutions obtained using SgDQE-LL and SgDQE-LH elements for $g/l_{x}=0.00001$ are compared with classical solutions. Due to non-availability of gradient plate solutions, comparison is made between SgDQE-LL and SgDQE-LH elements for higher $g/l_{x}$ values. The results for $g/l_{x}=0.00001$ agree well with the classical solutions and close proximity is seen in results obtained by SgDQE-LL and SgDQE-LH elements for higher $g/l_{x}$ values.

\setlength{\extrarowheight}{.5em}
\begin{table}[H]
    \centering
    \caption{Comparison of nondimensional central deflection $w_{p}(\frac{l_{x}}{2},\frac{l_{y}}{2})$ for a gradient plate under a central point load.}
      
    \begin{tabular}{p{3.0em}p{5em}p{3.75em}p{3.75em}p{3.75em}p{3.75em}}
     \\ \hline
support        & $g/l_{x}$      & 0.00001	         & 0.05           & 0.1     & 0.5    \\  \hline  
 	    & SgDQE-LL         &  0.7668  &  0.5472   & 0.3458  &  0.02867    \\
	CFCF  & SgDQE-LH       & 0.7518 &   0.5452 &   0.34409 & 0.02845 \\ 
  & DQM\cite{Wangb}
         & 0.7601 &  ------ &  ------ &  ------  \\ 
  & FEM\cite{Wangb}
         & 0.7648 &  ------ &  ------ &  ------  \\  \hline 
           
& SgDQE-LL      &     2.3373 &  2.2508 &   2.0459 &   0.6145  \\ 
  SFSF     & SgDQE-LH      &   2.3389 &   2.2527 &   2.0529 &   0.6158  \\ 
    & DQM\cite{Wangb}
         & 2.3172 &  ------ &  ------ &  ------  \\
  & FEM\cite{Wangb}
         & 2.3217 &  ------ &  ------ &  ------  \\   \hline 
  
  & SgDQE-LL      &   1.1591 &   0.8644  & 0.6266 &  0.0761  \\ 
  SFCF     & SgDQE-LH      &  1.1605  & 0.8596 &  0.5950  & 0.0664  \\ 
     & DQM\cite{Wangb}
         & 1.1519 &  ------ &  ------ &  ------                                           \\
   & FEM\cite{Wangb}
         & 1.1566 &  ------ &  ------ &  ------                                            \\ \hline  
   
        \end{tabular}
    \label{Behav_All_Disp_Plate_Ptl}
\end{table}

\subsubsection{Gradient Kirchhoff plate under cylindrical bending}

A gradient plate with two opposite sides clamped $(x=0,l_{x})$ subjected to udl and under cylindrical bending is analysed. The maximum deflection obtained using SgDQE-LL and SgDQE-LH elements with 15 grid points are compared with analytical solutions \cite{Besko3p} in Table \ref{Behav_Cycl_Bend_Plate_udl}. The results obtained using both elements are in excellent match with the analytical solutions for all $g/l_{x}$ values considered.
\setlength{\extrarowheight}{.5em}
\begin{table}[H]

    \centering
    \caption{Comparison of maximum nondimensional deflection for a gradient plate under cylindrical bending.}
      
    \begin{tabular}{p{4.3em}p{5em}p{3.75em}p{3.75em}p{3.75em}p{3.75em}}
      \\ \hline
support        & $g/l_{x}$      & 0.00001	         & 0.05           & 0.1     & 0.5 \\  \hline  
	Cylindrical 	    & SgDQE-LL         &  0.2604  &  0.1678   & 0.1028  &  0.0083   \\ 
bending  & SgDQE-LH   & 0.2629 &   0.1692 & 0.1028 &   0.0083 \\ 
 & Ref. \cite{Besko3p}
         & 0.2583 &  0.1678 & 0.1028 &  0.0083  \\  \hline 
  
        \end{tabular}
    \label{Behav_Cycl_Bend_Plate_udl}
\end{table}

\section{Conclusion}

A novel differential quadrature beam element was proposed to solve a sixth order partial differential equation associated with non-classical beam theories. This methodology was extended to formulate two new and different versions of differential quadrature plate elements for non-classical gradient elasticity theory. A new way to account for the non-classical boundary conditions associated with the gradient elastic beam and plate theories was introduced. The efficiency of the proposed elements was established through application to flexural problems of beams and plates.

\section*{Acknowledgement}

The authors would like to thank Professor Xinwei Wang, \textit{Nanjing University of Aeronautics and Astronautics, People’s Republic of China}, for the useful technical discussions.

\section*{APPENDIX}

\subsection*{Analytical solutions for static analysis of gradient elastic Euler-Bernoulli beam}

To obtain the static deflections of the gradient elastic Euler-Bernoulli beam which is governed by Equation \ref{eq:EOM_Beam}, we assume a solution of the form

\begin{align*}   \label{A1}     
w(x)=c_{1}{x}^{3}+c_{2}{x}^{2}+c_{3}{x}+c_{4}+c_{5}\,g^{4}\sinh(x/g)+c_{6}\,g^{4}\cosh(x/g)-(q_{b}/24EI)x^{4} \\ \tag{A1}
\end{align*}

\noindent The constants $c_{1}-c_{6}$ are determined with the aid of boundary conditions listed in Equation (\ref{eq:BC_Cl_Beam}) and (\ref{eq:BC_NCl_Beam}). After applying the boundary conditions the system of equations are expressed as:

\begin{align*}        
[K]\{\delta\}=\{f\} \tag{A2}
\end{align*}

here $K$ is the coefficient matrix, $f$ is the vector corresponding to the load and $\{\delta\}=\{c_{1},c_{2},c_{3},c_{4},c_{5},c_{6}\}$ is the unknown constant vector to be determined. Once the unknown constants are determined then the displacement solution is obtained from the Equation (A1). The slope and curvature at any point along the length of the beam can be obtained by performing the first and second derivatives of the deflection. The shear force, bending moment and higher moment are obtained by substituting the Equation (\ref{A1}) in Equations(\ref{eq:BC_Cl_Beam}) and (\ref{eq:BC_NCl_Beam}). The following are the expressions for support reactions: \\
\begin{align*}  
 Shear force : V=&6EI{c}_{1} - \frac{q_{b}x}{EI}  \\ \\
 Bending moment: M=&2EI({c}_{2} + 3{c}_{1}x) + \frac{q_{b}}{EI}\big[g^{2} -\frac{{x}^{2}}{2}\big]\\ \\
 Higher moment: \bar{M}=&6EI{g}^{2}c_{1} + EI{g}^{3}\text{cosh}(x/g){c}_{5} + EI{g}^{3}\text{sinh}(x/g)c_{6} - q_{b}{g}^{2}x\\
\end{align*}

The following are the list of simultaneous equations to determine the unknown constants for different boundary conditions: \\

\noindent (a) Simply supported beam :

$$[K]=
\begin{bmatrix}
 0 & 0 & 0 & 1 & 0 & g^4\\ \\
0 & 2 & 0 & 0 & 0 & 0 \\ \\
L^{3} & L^{2} & L & 1 & g^{4}\sinh(L/g) & g^{4}\cosh(L/g)\\ \\
6L & 2 & 0 & 0 & 0 & 0\\ \\
 0 & 2 & 0 & 0 & 0 & g^{2}\\ \\
6L & 2 & 0 & 0 & g^{2}\sinh(L/g) & g^{2}\cosh(L/g)\\ \\
\end{bmatrix}
$$

$${f}=
\begin{Bmatrix}
 0 \\ \\
g^{2}q_{b}/EI \\ \\
-q_{b}L^{4}/24EI\\ \\
g^{2}q_{b}/EI-q_{b}L^{2}/2EI\\ \\
 0 \\ \\
-q_{b}L^{2}/2EI\\ \\
\end{Bmatrix}
$$
 
\noindent (c) clamped beam :
$$[K]=
\begin{bmatrix}
 0 & 0 & 0 & 1 & 0 & g^4\\ \\
0 & 0 & 1 & 0 & g^{3} & 0 \\ \\
L^{3} & L^{2} & L & 1 & g^{4}\sinh(L/g) & g^{4}\cosh(L/g)\\ \\
3{L}^{2} & 2L & 1 & 0 & g^{3}\cosh(L/g) & g^{3}\sinh(L/g)\\ \\
 0 & 2 & 0 & 0 & 0 & g^{2}\\ \\
6L & 2 & 0 & 0 & g^{2}\sinh(L/g) & g^{2}\cosh(L/g)\\ \\
\end{bmatrix}
$$

$${f}=
\begin{Bmatrix}
 0 \\ \\
 0 \\ \\
-q_{b}L^{4}/24EI\\ \\
-q_{b}L^{3}/6EI\\ \\
 0 \\ \\
-q_{b}L^{2}/2EI\\ \\
\end{Bmatrix}
$$

\end{document}